\documentclass[12pt]{article}
\usepackage{amssymb,amsmath,amsthm,graphicx,ulem}

\pdfoutput=1

\usepackage{graphicx,subfigure}
\usepackage{epsfig}
\usepackage{amsmath}
\usepackage{amsfonts}
\usepackage{amssymb}
\usepackage[usenames]{color}
\usepackage[a4paper,left=2.cm,right=2.cm,top=2.5cm,bottom=2.5cm]{geometry}

\newcommand{\beq}{\begin{equation}}
\newcommand{\eeq}{\end{equation}}
\newcommand{\be}{\begin{equation}}
\newcommand{\ee}{\end{equation}}
\newcommand{\beqa}{\begin{eqnarray}}
\newcommand{\eeqa}{\end{eqnarray}}
\newcommand{\beqar}{\begin{eqnarray*}}
\newcommand{\eeqar}{\end{eqnarray*}}
\newcommand{\bea}{\begin{eqnarray}}
\newcommand{\eea}{\end{eqnarray}}

\newcommand{\ie}{{\it i.e.}\ }

\numberwithin{equation}{section}

\begin{document}

\allowdisplaybreaks

\normalem

\title{On the Nonlinear Stability of Asymptotically \\
Anti-de Sitter Solutions}
\vskip1cm
\author{\'Oscar J. C. Dias${}^{\,a}$, Gary T. Horowitz${}^{\,b}$, Don Marolf${}^{\,b}$,
Jorge E. Santos${}^{\,b}$\\
\\ ${}^{\,a}$ Institut de Physique Th\'eorique, CEA Saclay, \\ CNRS URA 2306, F-91191 Gif-sur-Yvette, France \\
 ${}^{\,b}$ Department of Physics, UCSB, Santa Barbara, CA 93106, USA \\ \\
 \small{oscar.dias@cea.fr, gary@physics.ucsb.edu, marolf@physics.ucsb.edu, jss55@physics.ucsb.edu}}

 \date{}

\maketitle

\begin{abstract}
\noindent Despite the recent evidence that anti-de Sitter spacetime is nonlinearly unstable, we argue that many asymptotically anti-de Sitter solutions are nonlinearly stable. This includes geons, boson stars, and black holes. As part of our argument, we calculate the frequencies of long-lived gravitational quasinormal modes  of AdS black holes in various dimensions. We also discuss a new class of asymptotically  anti-de Sitter solutions describing noncoalescing black hole binaries.
\end{abstract}

\newpage

%%%%%%%%%%%%%%%%%%%%%%%%%%%%%%%%%%%%%%%%%%%%%%%%%%%%%%%%%%%%%%%%%%%%%%%%%%%

\tableofcontents

%%%%%%%%%%%%%%%%%%%%%%%%%%%%%%%%%%%%%%%%%%%%%%%%%%%%%%%%%%%%%%%%%%%%%%%%%%%

\section{Introduction}
%%%%%%%%%%%%%%%%%%%%%%%%%%%%%%%%%%%%%%%%%%%%%%%%%%%%%%%%%%%%%%%%%%%%%%%%%%%

It has long been known that anti-de Sitter (AdS) spacetime is stable to linearized perturbations, and positive energy theorems ensure that AdS cannot decay. However, there is growing evidence that AdS is nonlinearly unstable, in the sense that any finite perturbation will eventually become large and, at least in many cases, form a small black hole \cite{Bizon:2011gg,Jalmuzna:2011qw,Dias:2011ss}. There were at least two earlier general arguments for this instability, which take the following form. Physically,  AdS is like a confining box. If Einstein's equations are sufficiently ergodic, then one might expect any generic finite perturbation to eventually explore all configurations consistent with the conserved quantities, including small black holes \cite{Anderson:2006ax}. A more mathematical argument is that the linearized perturbations do not decay. So nonlinear corrections (which usually involve integrals of powers of the linearized modes) are expected to become large \cite{Dafermos}.  Here, as is standard in the relativity community, our use of the term AdS refers to the universal covering space of the surface $\sum_{i=1}^{d-1}(X^i)^2 -T_1^2 -T_2^2 = -L^2$ in a Minkowski space of signature $(d-1,2)$.  This is what string theorists call ``global'' AdS, as opposed to just the Poincar\'e patch of AdS which is often used in studies of the AdS/CFT correspondence.

The recent evidence for the nonlinear instability of AdS came first from numerical evolution of a spherical scalar field coupled to gravity \cite{Bizon:2011gg,Jalmuzna:2011qw}. It was found that no matter how small one makes the amplitude of the scalar field, the energy in the scalar field cascades to smaller and smaller scales and eventually forms a black hole. The time to form the black hole $t_{BH}$ was found to scale with the initial amplitude $\epsilon$ via $t_{BH} \sim 1/\epsilon^2$.
The instability (and this time scale) can also be seen in terms of resonances\footnote{Resonances arise when there are nontrivial  linear relations among the frequencies of the first order equations.} in the perturbative construction of the solution \cite{Bizon:2011gg}.
For a generic initial perturbation, resonances appear at third order which require that one add higher frequency modes with linearly growing amplitude. In other words, the perturbative solutions $\phi$ have the schematic form

\begin{equation}
\label{exp}
\phi = \phi_0  +  A \phi_0^3 +  t B \phi_0^3 + ...
\end{equation}
where $\phi_0$ is a solution to the linearized problem and $A,B$ are appropriate operators\footnote{For example, the term $A \phi_0^3$ represents a a trilinear operator with `diagonal' arguments $\phi_0, \phi_0, \phi_0$.}.  For $t \sim 1/\phi_0^2$, the term linear in $t$ becomes comparable to the lowest order term. After this time one expects  nonlinearities to dominate. Since total energy is conserved, the energy is transferred to the higher frequency mode. Here it is important to note that, in describing the above result, we have already included the possibility to resum perturbation theory by shifting the mode frequencies at each order.  The expansion takes the form \eqref{exp} even after all possible such shifts have been implemented. One expects the transfer of energy to higher frequencies to continue, eventually forming a black hole.

The purely gravitational problem has a similar perturbative structure \cite{Dias:2011ss}.
%Gravitational normal mode frequencies are again integer multiples of the AdS frequency,
For a generic initial perturbation, resonances again arise at third order which transfer energy to higher frequency modes. Although the numerical evolution has not yet been performed for this case, it is natural to expect eventual black hole formation
%on a timescale $t_{BH} \sim 1/\epsilon^2$
in this context as well.

Although generic perturbations of AdS lead to a nonlinear instability, it was shown in
\cite{Dias:2011ss} that if one starts with certain individual graviton modes, nonlinear corrections can be added without triggering an instability. The resulting nonlinear analog of the linearized mode is called either a {\it geon}\footnote{This use of the term geon should should not be confused with the use of the term in, e.g., \cite{Friedman:1993ty}.} or an {\it oscillon} depending on whether or not it has a Killing field that is timelike in some region. Given the linearized mode, there is a one parameter family of either geons or oscillons labelled by their energy. They can be viewed as gravitational analogs of boson stars \cite{Liebling:2012fv}.

The two general arguments for the nonlinear instability of AdS mentioned above would seem to apply to geons and boson stars as well, suggesting they too will be nonlinearly unstable. We will argue that this is incorrect: Most asymptotically AdS solutions which are stable to linear perturbations appear to be  stable nonlinearly as well.  Strictly speaking, the lack of a useful Killing field means that our methods do not directly address the stability of oscillons.  However, we expect their behavior to be similar to that of geons and other solutions with some notion of time translation symmetry.

The key difference between pure AdS and these asymptotically AdS solutions is that in AdS, the normal mode frequencies are all integer multiples of the AdS frequency. This results in a large number of resonances. Other solutions with less symmetry do not have this property. However, they still have normal modes with very large angular momentum $\ell$ which are supported at large radius due to the centrifugal barrier. In this asymptotic region, the metric approaches AdS and the normal mode frequencies approach those of AdS. In other words there are approximate resonances at large $\ell$. Indeed, the purely gravitational instability of AdS seems to rely on these large $\ell$ modes. Since a generic perturbation always contains some large $\ell$ modes, one might wonder if the AdS instability carries over to these solutions as well.

\begin{figure}[t]
\centerline{\includegraphics[width=.50\textwidth]{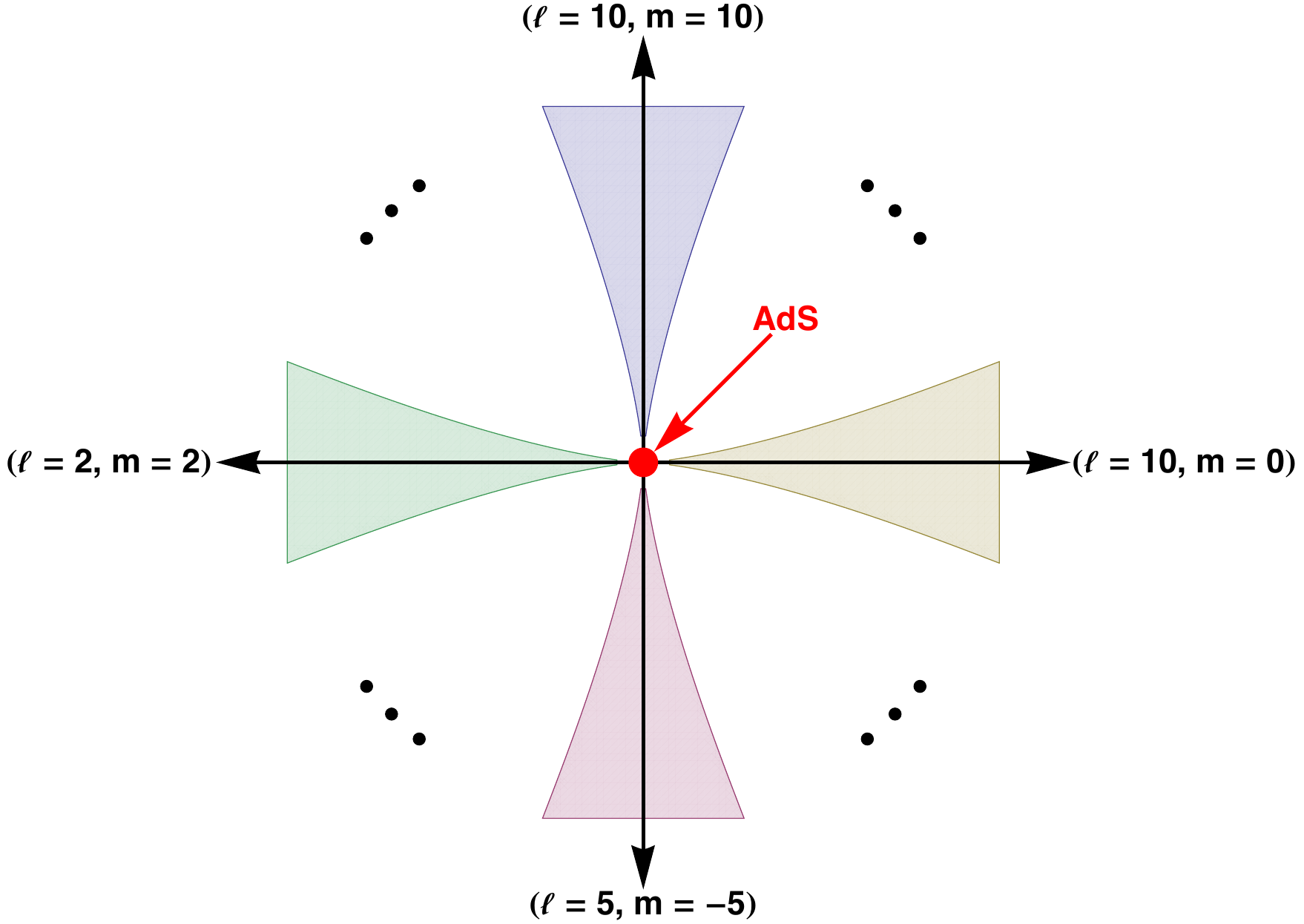}}
\caption{
The structure of solutions near empty AdS.  The shaded regions denote islands of stability near 1-parameter families of geons or oscillons, both indicated by black lines with arrows pointing toward increasing amplitude.  Because perturbation theory about empty AdS leads to geons only for a measure zero set of seed solutions, each such region has been drawn so that empty AdS lies at a cusp.} \label{fig:is}
\end{figure}

By studying the perturbative expansion in more detail, we will argue that these approximate resonances are not strong enough to trigger an instability\footnote{More precisely, we will argue that instabilities arise only in high dimensions for perturbations of low differentiability.}.
In particular, geons appear to be nonlinearly stable. Since geons are continuously connected to pure AdS by decreasing their amplitude to zero, one obtains an interesting picture of stability in the space of asymptotically AdS solutions. The geons radiate out from AdS, and are stable in finite neighborhoods whose size decreases to zero  as one approaches AdS (see Fig.~\ref{fig:is}). In other words, there are ``islands of stability" in a neighborhood of AdS. A similar picture of stability has recently been rigorously established for a simpler nonlinear PDE on a compact space: The nonlinear Schr\"odinger equation on a torus \cite{Colliander,Carles,Faou}.

Given the stability of general solutions without horizons,  AdS black holes that are linearly stable appear to be nonlinearly stable as well since the modes eventually decay (although the large $\ell$ modes live a very long time). However, the AdS boundary conditions do give rise to a novel class of black hole solutions: noncoalescing binaries. We will argue that there are asymptotically AdS solutions describing two black holes in a binary orbit which does not decay.
 The radiation produced by the black holes bounces off infinity and stops the orbit from shrinking, \ie the gravitational radiation is in a standing wave configuration. Although these noncoalescing orbits can be stable against radial perturbations, at least one black hole is unstable to superradiance. So eventually, the angular velocity is reduced and the black holes spiral together. Similar orbits around the Kerr black hole were proposed in \cite{Press:1972zz}, and recently realized in \cite{Cardoso:2011xi}, even though they are based on a different mechanism.
 
We begin in the next section with a brief discussion of the nonlinear Schr\"odinger equation on a torus. We will argue that the stability of asymptotically AdS solutions is similar to the structure found in this simpler system. To discuss the role of approximate resonances, we need to know the rate at which the normal modes with large angular momentum (or quasinormal modes for AdS black holes) approach those of AdS. This is computed in section 3 using a WKB approximation. We consider arbitrary dimension and both rotating and nonrotating black holes. Section 4 contains the main argument that perturbation theory converges despite the approximate resonances. Section 5 describes the construction of noncoalescing binaries. The final section discusses some generalizations of these results and their implications for gauge/gravity duality.

%%%%%%%%%%%%%%%%%%%%%%%%%%%%%%%%%%%%%%%%%%%%%%%%%%%%%%%%%
\section{Nonlinear Schr\"odinger equation on a torus}
%%%%%%%%%%%%%%%%%%%%%%%%%%%%%%%%%%%%%%%%%%%%%%%%%%%%%%%%%
\label{NLSE}

Consider the cubic nonlinear Schr\"odinger equation on a torus of dimension $D>1$\be\label{NLS}
i\partial_t u + \nabla^2 u = \lambda |u|^2 u.
\ee
This equation has a conserved energy
\be
E = \int d^Dx \left (\frac{1}{2} |\nabla u|^2 + {\lambda\over 4} |u|^4\right )
\ee
and a conserved norm
$N = \int d^Dx |u|^2$. For $\lambda \ge 0$, solutions do not blow up in finite time. Given smooth initial data $u(x, t=0)$, there is a smooth solution for all $t>0$.

 It has recently been shown that the solution $u=0$ is unstable in the sense that there are nearby solutions in which the energy  cascades to higher and higher Fourier modes\footnote{This doesn't happen in one spatial dimension since (\ref{NLS}) is integrable in that case.} \cite{Colliander,Carles}. This can be made precise  by looking at higher Sobelev norms defined as follows. Let $\hat u_n(t)$ be the $n^{th}$ Fourier mode of $u(x,t)$ on ${\mathbb T}^D$ and define
 \be
 || u(t)||^2_{H^s} = \sum_{n\in {\mathbb Z}^D} (1+n_jn^j)^{s} |\hat u_n(t)|^2.
 \ee
where $n_j$ are integers labeling the modes. It was shown that for any small number $\delta$ and large number $K$, there is a smooth solution for which $ || u(0)||^2_{H^s} < \delta$ and $ || u(T)||^2_{H^s} > K$ at some time $T$. This holds for any $s>1$. (The $s=1$ norm is constant due to energy conservation).

 This is directly analogous to the instability of AdS with the one exception that there are no analogs of black holes in this theory to cut off the cascade. The reason for the instability is also the same. Solutions of the linear equation ($\lambda = 0$) are of the form
\be\label{planewave}
 u(x,t) = A e^{i(n_j x^j - \omega t)}
 \ee
 with $\omega = n_j n^j$. Thus there are lots of resonances.

 A simple class of nonlinear solutions to (\ref{NLS})  are plane waves. These take exactly the same form as the linear solutions (\ref{planewave}) but now the frequency depends on the amplitude, $\omega = n_j n^j + \lambda A^2$. They are analogs of the geons in the gravitational problem. Faou, Gauckler and Lubich have recently shown that for almost all $A$, these nonlinear solutions are stable and there is no energy cascade \cite{Faou}.  More precisely, for $s>D/2$, the $H^s$ Sobolev norm remains small (at least for times that extend to arbitrary negative powers of the size of the perturbation). The proof relies on the fact that for almost all choices of $A$, most frequencies are now nonresonant. There remain a finite number of resonances, but they can only transfer energy among themselves. Since the energy cannot leak to high frequency modes, the higher Sobolev norms do not grow.

 The picture that emerges is that if a perturbation of $u=0$ has one mode with amplitude much larger than all the others, then the solution will stay close to the plane wave associated with that mode. Otherwise the modes compete and nonlinearities drive the energy into higher and higher frequency modes.
 We will argue that stability issues in AdS are similar.

%%%%%%%%%%%%%%%%%%%%%%%%%%%%%%%%%%%%%%%%%%%%
\section{Asymptotic behavior of quasinormal modes}\label{sec:testField}
%%%%%%%%%%%%%%%%%%%%%%%%%%%%%%%%%%%%%%%%%%%%%%%%%%%%%%%%%%%%%%%%%%%%%%
To study the nonlinear stability of general stationary backgrounds such as geons\footnote{We include geons here even though their Killing field is spacelike at infinity.}, boson stars, and black holes, we need to know how quickly the frequency of the large angular momentum perturbations approach those of AdS.  Since these modes are located at large radius, we may expect the leading deviation from the AdS frequencies to be determined by parameters like the total mass $M$ and the total angular momentum $J$ of the background. While the detailed form of these frequencies may also depend on higher moments of the energy-momentum distribution (which contribute to the metric at the same order in AdS), we may expect the discrepancies $\Delta \omega_{n} = \omega_{n} - \omega_{n}^{AdS}$ from the AdS values to decay at large angular momentum $\ell$ with the same power law for all values of these parameters\footnote{The imaginary part vanishes for any linearly stable horizon-free background.}.  This should be a consequence of the well-known Fefferman-Graham expansion for asymptotically AdS spacetimes\footnote{\label{sources} Backgrounds which differ from AdS at large $r$ by terms larger than those associated with $M,J$, etc. should have $\Delta \omega_n \sim \ell^{-\alpha}$ with smaller $\alpha$.  Since the approximate resonances are even weaker, they should be even more stable.}.

We test this hypothesis (and determine the associated power law) in two steps below, first using the global AdS$_d$ Schwarzschild black hole in section \ref{sec:WKB}, and then using a simple rotating black hole solution in section \ref{sec:WKBmp}.  In both cases we find  ${\rm Re} \ \Delta \omega_n \sim K_d \ell^{-\frac{d-3}{2}}$ at large $\ell$ where only $K_d$ depends on $M$ and $J$.  This analysis also displays the expected long-lived quasinormal modes first observed in the case of a massive scalar field in \cite{Festuccia:2008zx,Berti:2009wx}.  For completeness, we also treat such scalar field perturbations below.

%%%%%%%%%%%%%%%%%%%%%%%%
\subsection{Long-lived quasinormal modes: Schwarzschild-AdS black hole \label{sec:WKB}}
%%%%%%%%%%%%%%%%%%%%%%%%

We first briefly review the Kodama-Ishibashi harmonic decomposition of gravitational perturbations in the Schwarzschild-AdS black hole and discuss carefully the boundary conditions that preserve the global AdS asymptotics of the spacetime. Then we find the  quasinormal mode spectrum of modes with large angular momentum using a WKB approximation.

%%%%%%%%%%%%%%%%%%%%%%%%
\subsubsection{Perturbations of  Schwarzschild-AdS  and their boundary conditions \label{sec:BC}}
%%%%%%%%%%%%%%%%%%%%%%%%

 Consider the Schwarzschild-AdS$_d$ black hole, with horizon topology  $S^{d-2}$,   that asymptotes to global AdS$_d$,
\begin{equation}\label{gSch}
 ds^2=-fdt^2+\frac{dr^2}{f}+r^2d\Omega_{d-2}^2\,,\qquad \hbox{with} \quad f=1+\frac{r^2}{L^2}-\frac{r_m^{d-3}}{r^{d-3}}\,
\end{equation}
where $L$ is the AdS radius, $r_m$ is the mass parameter, $d\Omega_{d-2}^2$ is the line element of a unit radius $(d-2)$-sphere and the horizon is located at $r=r_+$ (the largest real root of $f$).

We want to study gravitational perturbations in this background that are constrained to obey the linearized Einstein equations subject to appropriate boundary conditions at the horizon and at the asymptotic boundary.
Since our $d$-dimensional background is locally the product of a 2-dimensional orbit spacetime (parametrized by the $\{t,r\}$ coordinates) and the base space $S^{d-2}$, we can decompose perturbations $h_{ab}$ according to how they transform under coordinate transformations on the sphere $S^{d-2}$. This  harmonic decomposition was worked out in detail by Kodama and Ishibashi \cite{Kodama:2003jz}. The most general perturbation of  \eqref{gSch} can be decomposed  into a superposition of three classes of modes: tensor, vector and scalar. This decomposition is particularly useful since it allows one to reduce the linearized Einstein equation associated to  perturbations of \eqref{gSch} to a set of three decoupled gauge invariant Kodama-Ishibashi (KI) master equations for the KI master fields $ \Phi^{(j)}_{\ell_j}$ which can be written in a compact form as,\footnote{Note that tensor perturbations exist only for $d\geq 5$ which we henceforward assume to be the case when discussing this sector.}
\begin{equation}
 \left(  \Box_2 - \frac{U_j}{f} \right) \Phi^{(j)}_{\ell_j}(t,r)=0 \,, \qquad  \hbox{where} \quad j=\{T,V,S\},
\label{KImaster}
\end{equation}
for tensor, vector, and scalar perturbations, respectively. Here $\Box_2$ is the d'Alembertian operator in the 2-dimensional orbit spacetime, and the potentials $U_j$ are further discussed below.  They depend on the properties of the background, namely on the mass parameter $r_m$ and cosmological length $L$, and on the eigenvalues of the associated (regular) tensor, vector, and scalar harmonics. The latter are, respectively, given by
\begin{equation}
 \lambda_j=\ell_j(\ell_j+d-3)- c_j \,, \qquad   \hbox{where} \quad \{(\ell_j,c_j)\}=\{ (\ell_T,2),(\ell_V,1), (\ell_S,0)\}
\label{KIeigenvalues}
\end{equation}
with integer $\ell_j$ obeying $\ell_T\geq 2$,  $\ell_V\geq 1$, and  $\ell_S\geq 0$.

We also study perturbations of an axisymmetric massive scalar field  $\Psi$ in the Schwarzschild-AdS$_d$ black hole background. This field obeys the Klein-Gordon equation,
\begin{equation} \label{KGeq}
 \left(  \Box_d - \mu^2 \right) \Psi (t,r,{\bf x}) = 0\,, \qquad \Psi(t,r,{\bf x})= r^{-\frac{d-2}{2}}\,\sum_{\ell_\psi}\Phi_{\ell_\psi}^{(\psi)}(t,r) \,S_{\ell_\psi 0}({\bf x})\,,
\end{equation}
where $S_{\ell_\psi 0}({\bf x})$ are the scalar spherical harmonics on the $(d-2)-$sphere with eigenvalues $\lambda_S$ defined in \eqref{KIeigenvalues}. Note that the spectrum of $\ell_\psi$ coincides with that of $ \ell_S$ and, because the background space-time is spherically symmetric, we can restrict our attention to axisymmetric modes without loss of generality.

It is useful to introduce the tortoise coordinate
\begin{equation}\label{SchwTartoise}
z=\int_r^\infty \frac{dr}{f}\,, \qquad 0\leq z< \infty\,,
\end{equation}
that  increases monotonically as we move from the asymptotic boundary, where  $z\sim L^2/r \to 0$, to the horizon where $z\sim -\frac{\ln(r-r_+)}{f^{\prime}(r_+)}\to +\infty$.
Since our background is  time-translation invariant,  the fields can be Fourier decomposed in time as $ \Phi^{(j)}_{\ell_j}(t,r)= e^{-i\omega_j t} \Phi^{(j)}_{\omega_j,\,\ell_j} (r)$.

In these conditions, the angular and time dependence of our gravitational and scalar field perturbations factors out, and studying equation \eqref{KImaster}  and \eqref{KGeq} boils down to solving the radial Schr\"odinger  equation,
\begin{equation}\label{SchroEq}
 \partial_z^2\Phi^{(j)}_{\omega_j,\,\ell_j}+\left(\omega_j^2-U_j \right)\Phi^{(j)}_{\omega_j,\,\ell_j}=0\,, \qquad  \hbox{with} \quad j=\{T,V,S,\psi\},
\end{equation}
for  tensor ($T$), vector ($V$), and scalar ($S$) gravitational perturbations and scalar field perturbation ($\psi$). The exact expression for the potentials $\{U_T,U_V,U_S,U_\psi\}$ can be found in equations (5.6), (5.15),  (3.2)-(3.8) of  KI  \cite{Kodama:2003jz}, and (3.6) of \cite{Festuccia:2008zx}, respectively.

Solutions of this equation have to obey appropriate boundary conditions.  At the horizon, only ingoing modes are allowed. To enforce this condition, we change to ingoing Eddington-Finkelstein coordinates (appropriate to extending the analysis through the horizon) and demand regularity in this coordinate system.  The metric of the Schwarzschild-AdS$_d$ black hole asymptotically approaches that of global AdS$_d$; we would like to restrict to perturbations that preserve this behavior. The KI master variables have the asymptotic expansion,
 \begin{equation}\label{KIasymp}
\Phi^{(j)}_{\omega_j,\,\ell_j}{\bigl |}_{z\to 0}  \sim \Phi_0 \,r^{\frac{d-6}{2}+a_j}+\Phi_1\, r^{-\frac{d-4}{2}-a_j}  \,, \qquad  \hbox{with} \quad a_T=2,\, a_V=1,\, a_S=0\,.
\end{equation}
 The linear differential map $h_{ab}^{(j)}=h_{ab}^{(j)}\left( \Phi^{(j)} \right)$ that reconstructs the metric perturbations (in a given gauge) can be found in  \cite{Kodama:2003jz}. The requirement that these metric perturbations are asymptotically global AdS$_d$ in the sense of \cite{Boucher:1983cv,Henneaux:1985tv}, imposes the conditions \cite{Michalogiorgakis:2006jc,Dias:2011ss}:
   \begin{equation}\label{KI:BC}
\Phi_1 =-\frac{3 r_m}{\lambda_S-2} \Phi_0, \quad \hbox{if  \, $j=S$  in  $d=4$}\,;    \qquad \Phi_0=0 , \quad \hbox{otherwise}\,.
\end{equation}
These are also natural boundary conditions in the context of AdS/CFT: they allow a non-zero expectation value for the CFT stress-energy tensor while keeping fixed the boundary metric.  Other boundary conditions that might be called asymptotically globally AdS were studied in \cite{Compere:2008us}, but turn out to lead to ghosts (modes with negative kinetic energy) and thus make the energy unbounded below \cite{Andrade:2011dg}.

In the case of the massive scalar field, the asymptotic expansion reads \cite{Breitenlohner:1982jf,Mezincescu:1984ev}
\begin{equation}
\label{AdSdecay}
r^{-\frac{d-2}{2}} \Phi^{(\psi)}_{\omega_\psi,\,\ell_\psi}{\bigl |}_{r\to +\infty}\sim \frac{\Phi_0}{r^{\Delta_-}}+ \frac{\Phi_1}{r^{\Delta_+}}\,, \qquad
 \hbox{where} \quad \Delta_\pm=\frac{d-1}{2}\pm \nu_\psi\,,\qquad \nu_\psi=
 \sqrt{\frac{(d-1)^2}{4}+\mu^2L^2}.
\end{equation}
 Stability of the AdS background (\ie scalar energy bounded below) requires the mass of the scalar field to obey the Breitenl\"ohner-Freedman (BF) bound $\mu^2 \ge \mu^2{\bigr |}_{BF}$, where $\mu^2{\bigr |}_{BF}= -(d-1)^2/(4L^2)$.  Furthermore, to have a well-defined initial value problem we must impose appropriate boundary conditions at $r \rightarrow \infty$.
 For scalars with mass in the range $\mu^2{\bigr |}_{BF} \le \mu^2 < \mu^2{\bigr |}_{BF} + 1/L^2$, there is a choice of boundary conditions that respect the full AdS isometries: one can impose either that $\Phi^{(\psi)}$ decays as $r^{-\Delta_+}$, or that it decays as $r^{-\Delta_-}$ (in this case with no term at order $r^{-\Delta_+}$). According to the AdS/CFT correspondence, this choice dictates whether the operator dual to $\Phi^{(\psi)}$ has dimension $\Delta_+$ or $\Delta_-$. We will consider both cases. For $\mu^2 \ge  \mu^2{\bigr |}_{BF} + 1/L^2$, only fast fall-off boundary condition $\Phi = {\cal O}(r^{-\Delta_+})$ leads to a theory with energies bounded below; the alternate boundary condition leads to ghosts with negative kinetic energies \cite{Andrade:2011dg}.    We thus choose  the boundary condition $\Phi_0=0$ in this case.

%%%%%%%%%%%%%%%%%%%%%%%
\subsubsection{WKB  analysis of quasinormal modes}
%%%%%%%%%%%%%%%%%%%%%%%%%%%

\begin{figure}[t]
\centerline{\includegraphics[width=.50\textwidth]{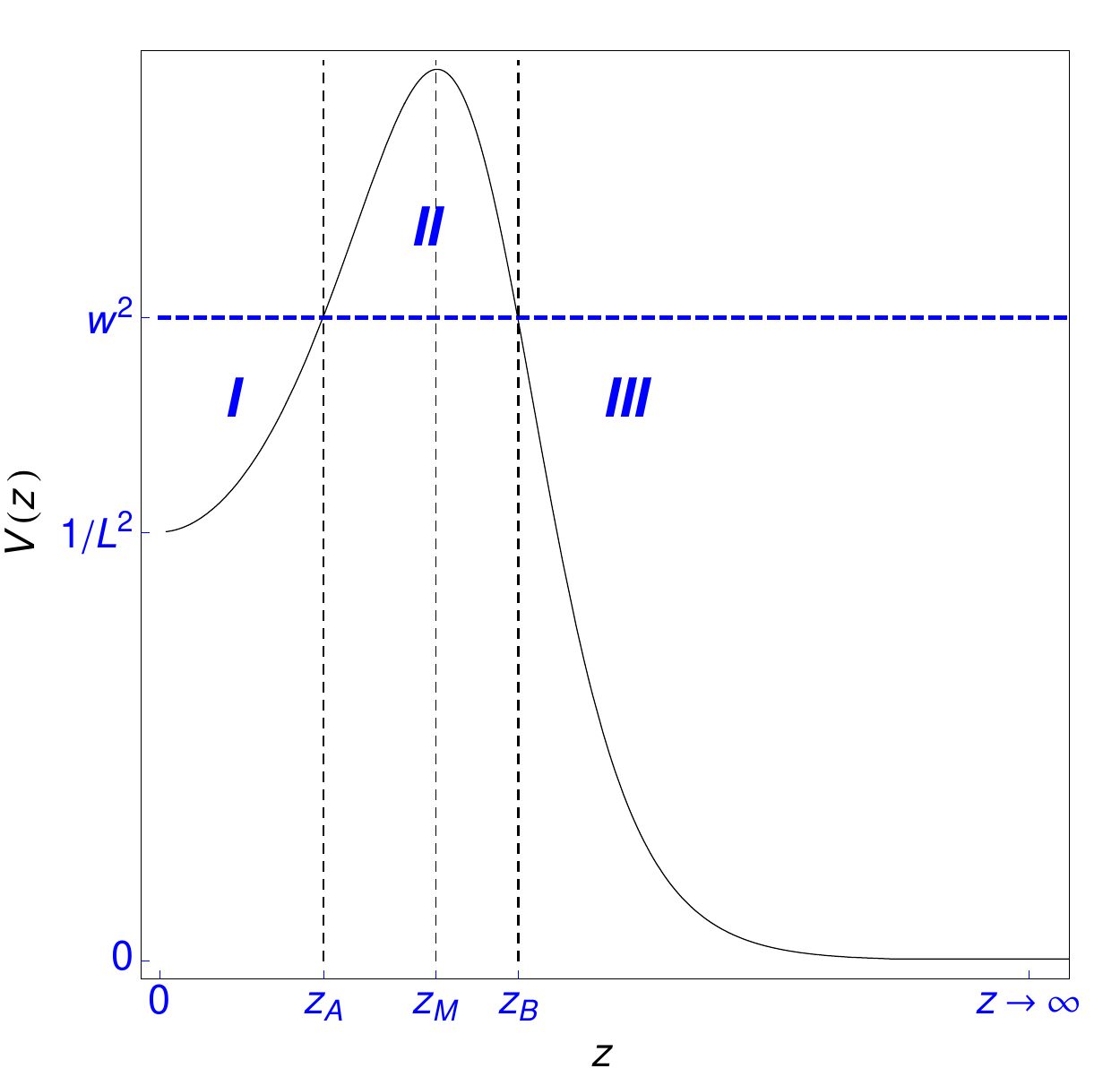}}
\caption{
WKB potential $V(z)$ for $r_+ = L$ and WKB allowed ($I,III$) and forbidden ($II$) regions for the corresponding particle orbits.
} \label{fig:wkb}
\end{figure}

We are interested in the large angular momentum WKB limit, $\ell_j\to\infty$, of the radial Schr\"oedinger  equation \eqref{SchroEq}. Define
\begin{equation}\label{Swkb}
\Phi^{(j)}_{\omega_j,\,\ell_j}(z)=e^{p_j\, S_{w_j,\,p_j}(z)} \,,\qquad  \omega_j=p_j\,w_j\,, \qquad p_j=\ell_j+\frac{d-3}{2}\,,
\end{equation}
so that \eqref{SchroEq} takes the form
\begin{equation}\label{SchroEqWKB}
\frac{1}{p} S_{w_j,\,p_j}''(z)+S_{w_j,\,p_j}'(z)^2-\left[V-w_j^2\right]=\frac{1}{p_j^2}\,\chi_j(z)\,,\qquad \hbox{where} \quad V=\frac{f }{ r^2}.
 \end{equation}
Note that $V$ is independent of the perturbation sector we look at, whereas $\chi_j(z)$ depends on the particular type of perturbation. We can solve this equation in a WKB approximation, which we review below, with expansion parameter $p_j^{-1}\ll 1$.
Introducing the wavefunction expansion $S_{w_j,\, p_j}=\sum_{k=0}^\infty p_j^{-k}\,\tilde S_{j,\,k} $ we find the leading order terms,
\begin{eqnarray}\label{SchwKGwkb}
 &&\tilde S_{j,\,0}(z)= \pm \,i \int^z\sqrt{w_j^2-V(x)}\,dx\,,\qquad \tilde S_{j,\,1}(z)=-\frac{1}{4} \ln \left[w_j^2-V(z)\right]\,,\\
 &&\tilde  S_{j,\,2}(z)= \pm \,\frac{i}{8}\int^z\left[ \frac{V''(x)}{ \left[w_j^2-V(x)\right]^{3/2} }+\frac{5}{4}\,\frac{V'(x)^2}{\left[w_j^2-V(x)\right]^{5/2}}-\frac{ 4 \chi_j (x)}{ \sqrt{w_j^2-V(x)}}\right] dx\,. \nonumber
\end{eqnarray}
These leading terms are enough to illustrate a relevant property of this WKB system:
the source term $\chi_j(z)$ only contributes at second or higher order of the expansion, \ie to $\tilde S_{j,\,k}(z)$ with $k\geq 2$. For our purposes it will be enough to include only the next-to-leading order correction,\ie $k\leq 1$. Thus we only need information concerning the potential $V(z)$. A typical behaviour for this potential is shown in Figure \ref{fig:wkb}. At first sight one might worry that all information about the specific sector of perturbations is lost at this WKB leading order. This is however not the case: this information will be encoded in the asymptotic boundary condition; see \eqref{wkb1z0} below.

We are interested in long-lived quasinormal modes. These are bound states with frequencies $L^{-2}<w_j^2<V(z_M)$ related in the WKB limit to particle orbits trapped in the exterior region limited by the  asymptotic AdS boundary ($z=0$) and the potential barrier at $z=z_A$ (see Figure \ref{fig:wkb}). However, as solutions to a wave equation they can slowly leak through the potential barrier to the black hole horizon at $z\to\infty$. The role of the large angular momentum $\ell$ is therefore to provide the potential barrier that, as we will find, traps the waves for sufficiently long time  to make these quasinormal modes long-lived before being absorbed by the  horizon. In a WKB analysis we have to consider three regions: the allowed region $I$ ($0<z<z_A$), the forbidden region $II$ ($z_A<z<z_B$), and a second allowed region $III$ ($z_B<z<\infty$), where $z_A, z_B$ are the turning points defined in Figure \ref{fig:wkb}.  Here the terms allowed, forbidden, and turning point refer to the particle orbits associated with the given modes in the WKB limit.

We want to find the quasinormal modes frequencies in the limit of large $p_j$ using a standard WKB approximation. We start at the asymptotic boundary and move downwards toward the horizon.
The potential $U_j$ in  \eqref{SchroEq} typically contains terms proportional to $p_j^2$ but also terms proportional to $r^2$ (that are inherited from the $r^2/L^2$ factor in the metric function $f$). At the conformal boundary, $z\to  0$, these terms compete with each other and both contribute to the asymptotic solution in region $I$. Our task is to find the associated asymptotic solution of the Schr\"oedinger  equation that obeys the asymptotic boundary condition \eqref{KI:BC}. More concretely, the asymptotic potential and solution obeying the desired boundary condition are  (as observed earlier, note that these do depend on the particular sector of perturbations through $\nu_j$)
\begin{equation}\label{wkb1z0}
U_j{\bigl |}_{ z \to  0}\sim \frac{p_j^2}{L ^2}+\left(4\nu _j{}^2-1\right)\frac{r^2}{4\, L ^4}\,, \qquad \quad
\Phi _{w_j,\, p_j}^{(j)}{\bigl |}_{ z \to  0}\sim \sqrt{z}\, J_{\nu_j}\left(p_j\,z\, \sqrt{w_j^2-\frac{1}{L^2}}\right),
\end{equation}
where   $J_{\nu_j}(x)$ is a Bessel function of the first kind and we defined
\begin{equation}\label{def:nu}
\nu_T=\frac{d-1}{2}\,,  \qquad  \nu_V=\frac{d-3}{2}\,,  \qquad  \nu_S=\frac{d-5}{2}\,,  \qquad
\nu_\psi=
 \sqrt{\frac{(d-1)^2}{4}+\mu^2L^2}\,.
 \end{equation}
The solution for $p_j\to\infty$ with finite $z\neq 0$ can be obtained by expanding \eqref{wkb1z0} for large $p_j z$  while  consistently promoting $z\,\sqrt{w_j^2-\frac{1}{L^2}}\equiv z\,\sqrt{w_j^2-V|_{z\to 0}}\to \int_0^{z} dx \sqrt{w_j^2-V(x)}$. This gives the desired WKB solution in region $I$,
\begin{equation}\label{wkb1zfinite}
\Phi _{w_j\,p_j}^{(j)\: I} \simeq \frac{2C^{(j)}_1}{\sqrt{p}}\left(w_j^2-V\right)^{-1/4}\cos \left[p_j \int_0^{z} \! \! \sqrt{w_j^2-V}dx  -\frac{\pi}{4}  (1+2 \nu_j) \right].
\end{equation}
Defining
\begin{equation}  \label{WKBgamma}
\eta_j \equiv p_j \int_0^{z_A} \! \! \! \! \sqrt{Q_j}\, dx  -\frac{\pi}{4}  (1+2 \nu_j)\,,\qquad\quad Q_j(z)\equiv w_j^2 -V(z)\,,
\end{equation}
and
\begin{equation}\label{wkb1parameters}
C^{(j)}_2\equiv C^{(j)}_1e^{i\,\eta_j }\,,\qquad C^{(j)}_3\equiv C^{(j)}_1e^{-i\,\eta_j }\,,
\end{equation}
we can rewrite  \eqref{wkb1zfinite} in the standard WKB form for region $I$ in which the corresponding particle orbits are allowed,
\begin{equation}\label{wkb1}
\Phi _{w_j,\,p_j}^{(j)\: I} \simeq  \frac{C^{(j)}_2}{p_j^{1/2} Q_j^{1/4}} \,{\rm exp}\left[ i\,p_j \int_{z_A}^{z}  \! \! \sqrt{Q_j}\, dx \right] +  \frac{C^{(j)}_3}{p_j^{1/2} Q_j^{1/4}} \,{\rm exp}\left[ -i \,p_j \int_{z_A}^{z}  \! \! \sqrt{Q_j}\, dx \right]  .
\end{equation}
Note that \eqref{wkb1parameters} encodes the asymptotic boundary condition information.
 Since region $II$  ($z_A<z<z_B$) is a forbidden region for the corresponding particle orbits, the WKB solution is
 \begin{equation}\label{wkb2}
\Phi _{w_j,\, p_j}^{(j)\: II}  \simeq
 \frac{C^{(j)}_4}{p_j^{1/2}|Q_j|^{1/4}}\,
 {\rm exp}\left[ -p_j\int_{z_A}^{z}  \! \!  \sqrt{|Q_j|}\,dx \right] +
 \frac{C^{(j)}_5}{p_j^{1/2}|Q_j|^{1/4}}\, {\rm exp}\left[
p_j\int_{z_A}^{z}  \! \! \sqrt{|Q_j|}\,dx \right],
 \end{equation}
where the respective amplitudes are fixed by the standard WKB connection formulas around the turning point $z=z_A$,
 \begin{equation}\label{wkb2parameters}
C^{(j)}_4= \frac{1}{2} \left (C^{(j)}_2 \, e^{-i\, \pi/4}+C^{(j)}_3 \,e^{i\, \pi/4}\right)\,,
\qquad C^{(j)}_5 =i \left (C^{(j)}_2 \,e^{-i\, \pi/4}-C^{(j)}_3 \,e^{i\, \pi/4}\right)\,.
\end{equation}

Finally,  region $III$ ($z_B<z<\infty$) is again an allowed region where the horizon is encountered  at $z\to \infty$. Its WKB wavefunction is
  \begin{equation}\label{wkb3}
\Phi _{w_j,\,p_j}^{(j)\: III}  \simeq   \frac{C^{(j)}_6}{p_j^{1/2}Q_j^{1/4}} \,{\rm exp}\left[
i\,p_j\int_{z_B}^{z} \sqrt{Q_j}\,dx \right]
+\frac{C^{(j)}_7}{p_j^{1/2}Q_j^{1/4}}\, {\rm exp}\left[
-i\,p_j\int_{z_B}^{z} \sqrt{Q_j}\,dx \right]\,.
 \end{equation}
The WKB connection formulas around the classical turning point $z=z_B$ determine the amplitudes of region $III$ as
\begin{equation}\label{wkb3parameters}
C^{(j)}_6 =  \left ( \frac{i \,C^{(j)}_4}{2\Gamma_j}+ C^{(j)}_5 \Gamma_j \right )e^{-i\,
\pi/4}\,,\qquad C^{(j)}_7 = \left ( -\frac{i \,C^{(j)}_4}{2\Gamma_j}+ C^{(j)}_5 \Gamma_j \right
)e^{i\,\pi/4}\,,
\end{equation}
where we introduced the quantity
\begin{equation}  \label{WKBbeta}
\ln \Gamma_j \equiv
p_j \int_{z_A}^{z_B}\!\! \sqrt{|Q_j|}\,dx\,.
\end{equation}
At this point, we impose the second boundary condition of our problem. At the horizon, regularity of the perturbations in the ingoing Eddington-Finkelstein coordinates allows only ingoing modes. This requires setting $C^{(j)}_7=0$ in \eqref{wkb3parameters}.

From the relations  \eqref{wkb1parameters},  \eqref{wkb2parameters}, and \eqref{wkb3parameters}  (with the boundary condition $C^{(j)}_7=0$) we find that quasinormal modes must obey the condition
\begin{eqnarray}
i(4\,\Gamma_j^2-1)\,e^{i \,\eta_j }+(4\,\Gamma_j^2+1)\,e^{-i \,\eta_j }=0 \qquad \Leftrightarrow \qquad   e^{-i \,\eta_j }+i\,e^{i\,\eta_j }\simeq 0\,, \quad \hbox{for}\:\:p_j \gg 1 \,,
 \label{QNcondition}
\end{eqnarray}
where $\eta_j $ and $\Gamma_j$ are defined in (\ref{WKBgamma}) and
(\ref{WKBbeta}), respectively, and in the last equality we used that $\Gamma_j^2\gg 1$ in the large $p_j$ limit of our WKB approximation $-$ see \eqref{WKBbeta}\footnote{Large $\Gamma_j^2$ corresponds to a small imaginary part of the frequency, and in the end of our computation we confirm that indeed one has $\Gamma_j^2\gg 1$: see \eqref{damptime}.}. The quasinormal modes are thus described by solutions of \eqref{QNcondition}.
That is, the real part of the frequencies, $w_{j,\,n}$,  of  the quasinormal mode spectrum  are quantized by the condition
\begin{equation}
\eta_j  (w_j) \equiv n\,\pi +\frac{\pi}{4} \qquad  \Leftrightarrow \qquad  p_j \int_0^{z_A} \! \! \! \! \sqrt{w_{j,\,n}^2-V}\, dx  =\frac{\pi}{2} (2n+1+\nu_j)\,, \quad n=0,1,2,\cdots
 \label{ressonance freq}
\end{equation}
with the non-negative integer $n$ being the overtone that fixes the number of radial zeros of the mode, and the factor $\nu_j$ that depends on the specific perturbation sector is defined in \eqref{def:nu}.
At this point we just need to find the damping timescale of our system, as determined  by the
imaginary part of the quasinormal mode frequencies. Dissipation occurs because the bound states of region $I$ can leak through the potential barrier of region $II$ and be absorbed by the black hole horizon. This is thus a standard WKB tunneling process where the damping timescale $\tau_j=({\rm Im}\, w_j)^{-1}$ is proportional to the inverse of the WKB transmission coefficient through the barrier, $\tau_j\sim \Gamma_j^2$, \ie
\begin{equation}
\tau_{j,\,n} \sim \exp \left( 2 p_j \int_{z_A}^{z_B}\!\! \sqrt{V-w_{j,\,n}^2}\,dx\right), \quad n=0,1,2,\cdots
 \label{damptime}
\end{equation}
where a prefactor can be explicitly computed but is irrelevant for our proposes since it is a finite, order 1, value. The massive scalar quasinormal mode spectrum in a large $\ell_S$ WKB approximation was first computed in  \cite{Festuccia:2008zx} (using an equivalent anti-Stoke line WKB formalism) and this WKB expression has been explicitly compared against a numerical computation in  \cite{Berti:2009wx}. For large angular momentum $\ell_j$, the damping timescale \eqref{damptime}  scales as $\tau_{j,\,n} \sim e^{\,\ell_j}$. The large $\ell_j$ quasinormal modes are thus  very long-lived.

\begin{figure}[t]
\centerline{\includegraphics[width=.50\textwidth]{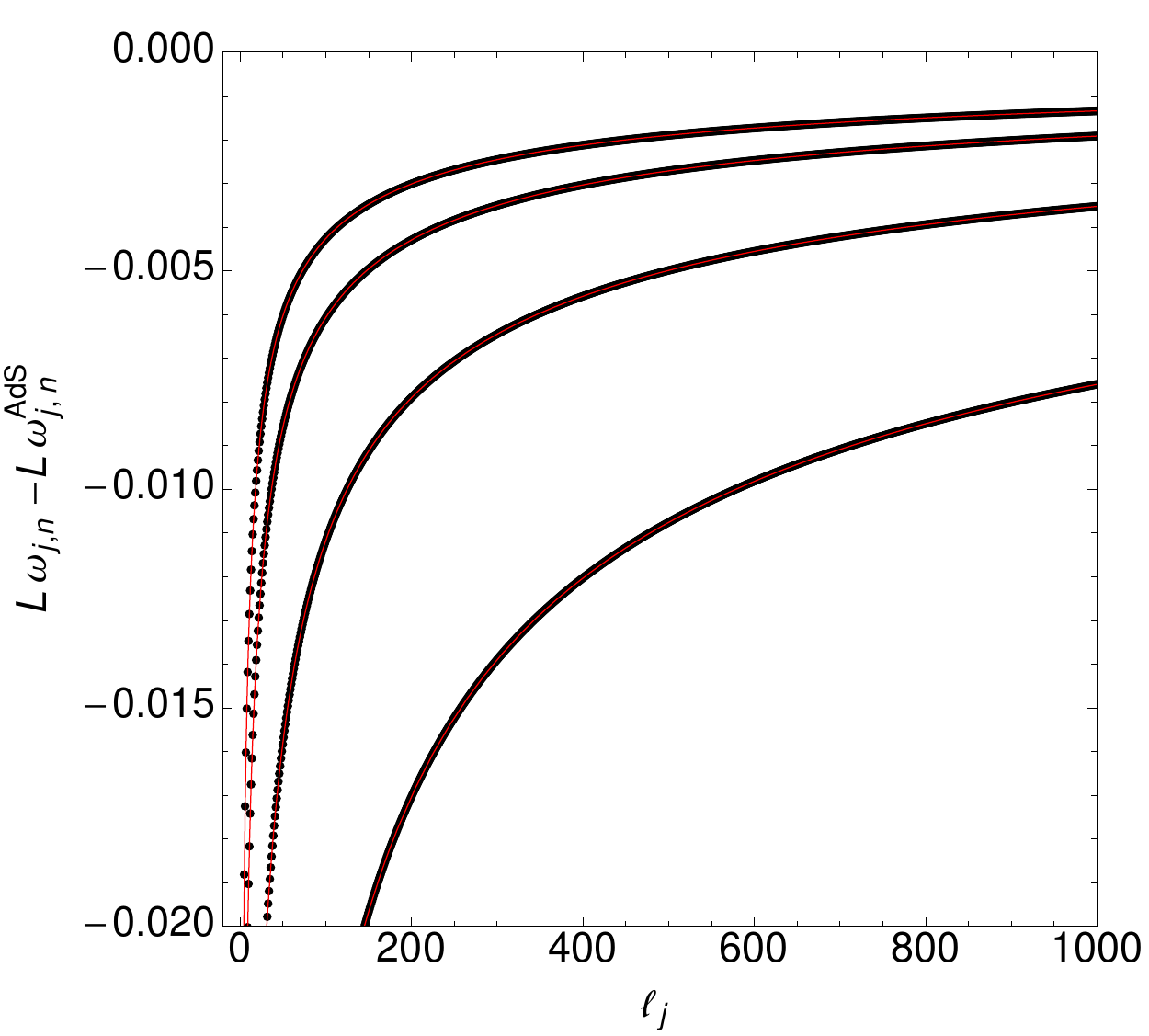}
\hspace{0.5cm}\includegraphics[width=.50\textwidth]{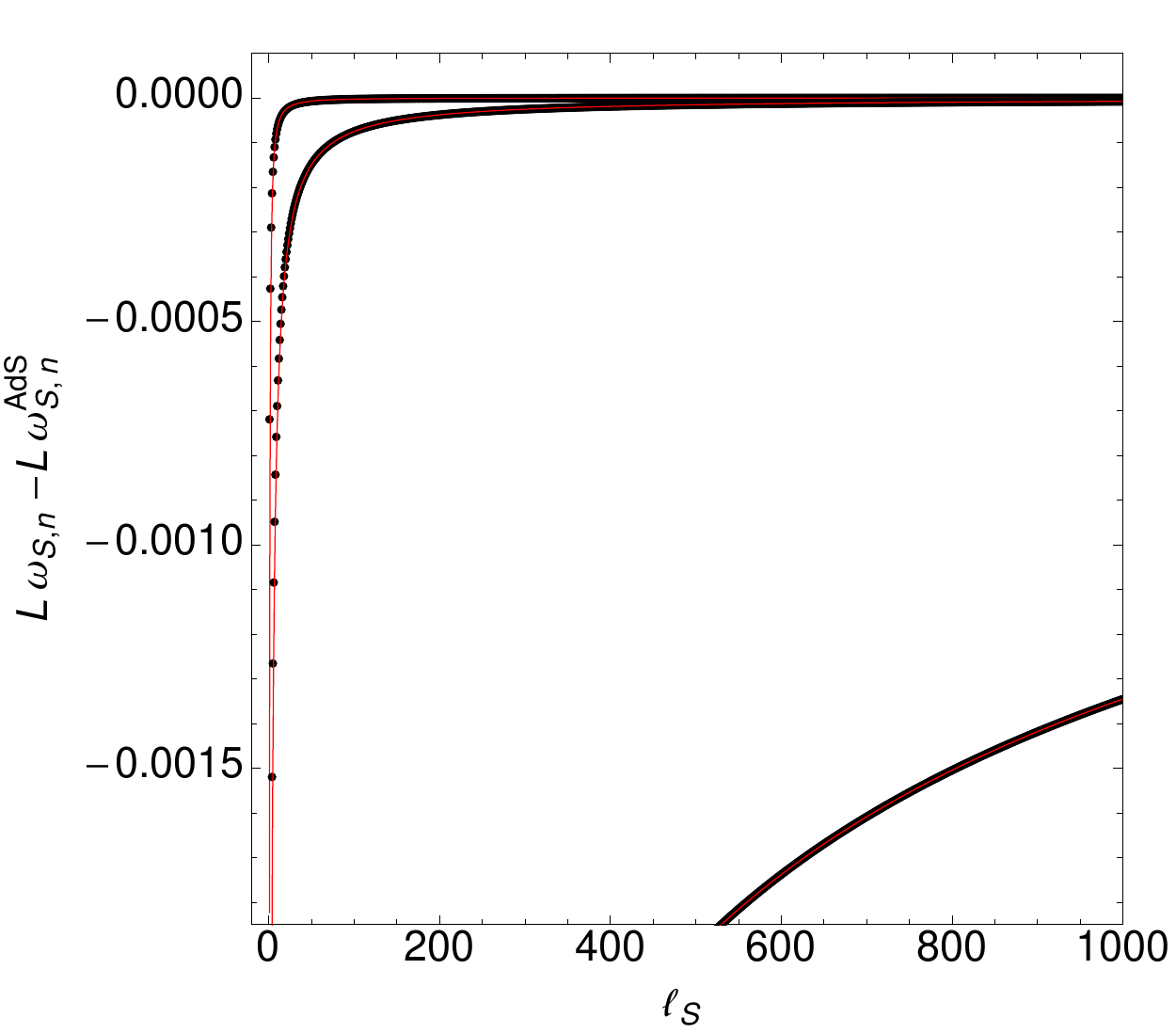}}
\caption{
{\it Left Panel:} Real part of the WKB quasinormal mode frequencies of the Schwarzschild-AdS black hole with respect to the AdS normal mode frequencies as a function of the WKB parameter $\ell_j$, in $d=4$ (and overtone $n=0$, and horizon radius $r_+=0.1 L$). Moving from left/top to right/bottom, the curves describe the scalar gravitational $(S)$, scalar field with BF mass $\mu^2L^2=-9/4$, vector gravitational $(V)$ and massless scalar field ($\mu=0$) cases (there are no regular tensor modes in $d=4$).
{\it Right Panel:} Evolution of $L \,\Delta \omega_{S,\,n} =L \,\omega_{S,\,n}-L \,\omega_{S,\,n}^{AdS}$ as the dimension $d$ increases, for the scalar gravitational $(S)$ case. From the left/top to right/bottom of the figure we have the lines corresponding to the cases $d=6$, $d=5$, and $d=4$.  In these plots the red continuous line is the best fit curve of the data to the curve $\Delta \omega_{j,\,n} = \alpha\, \ell_j^{\,\beta}$.  The best fit value of $\beta$ is $-(d-3)/2$ to machine precision.
} \label{fig:freq}
\end{figure}

We can solve \eqref{ressonance freq} and \eqref{damptime} to get valuable physical information.
As a first approximation, for frequencies $w_j\,L \sim  1$, the bound states are localized near the boundary and we can approximate the potential well around $z=0$ by a simple parabola $V\simeq \frac{1}{L ^2}\left(1+\frac{z^2}{L ^2}\right)$, with turning point at $z_A\simeq L  \sqrt{w_j^2 L ^2-1}$. In this regime the analysis is essentially blind to the black hole presence and captures only the global AdS asymptotics. Inserting these relations into \eqref {ressonance freq} we find that in this approximation the real part of the frequency, $\omega_n=w_{j,\,n}\, p_j$, naturally gives the normal mode frequencies of global AdS$_d$, for the three sectors of  gravitational modes and for the massive scalar field,
\begin{equation}
L\,\omega_{j,\,n}\simeq \left\{
\begin{array}{ll}
L\,\omega_{T,\,n}^{AdS}= (d-1)+\ell_T+2n\,, & \qquad \hbox{for tensor gravitational modes} \,,\label{wAdS} \\
L\,\omega_{V,\,n}^{AdS}= (d-2)+\ell_V+2n\,, & \qquad \hbox{for vector gravitational modes} \,, \\
L\,\omega_{S,\,n}^{AdS}= (d-3)+\ell_S+2n\,, & \qquad \hbox{for scalar gravitational modes} \,, \\
L\,\omega_{\psi,\,n}^{AdS}= \Delta_{\pm}+ \ell_S +2 n\,, & \qquad \hbox{for massive scalar modes} \,,
\end{array}
\right.
\end{equation}
for   $n=0,1,2,\cdots$. In the last case, the choice of $\Delta_+$ or  $\Delta_-$ (defined in (\ref{AdSdecay})) depends on the boundary condition for the scalar field. These expressions for the normal mode frequencies of AdS are known to be exact, \ie valid for any value of angular momentum $\ell_j$, in spite of the approximations used to obtain them.

Beyond this global AdS (parabola) approximation, the real and imaginary parts of the quasinormal mode frequencies can be computed by  numerically solving \eqref{ressonance freq} and \eqref{damptime}. We are particularly interested in the difference $\Delta \omega_{j,\,n} =\omega_{j,\,n}-\omega_{j,\,n}^{AdS}$. We find that in this  large $\ell_j$ WKB limit, the real part of the Schwarzschild-AdS quasinormal mode frequencies has the universal power law behavior
\begin{equation}\label{wSchAdS}
\omega_{j,\,n} \sim \omega_{j,\,n}^{AdS}+K_d^{(j)}\, \ell_j^{-\frac{ d-3}{2}}\,,
\end{equation}
where $K_d^{(j)}$ is a constant that depends on the kind of perturbation (and radial overtone), dimension $d$ and black hole radius in AdS units.  The exponent $-(d-3)/2$ was obtained by fitting $\Delta \omega_{j,\,n}$ to a power law for large $\ell_j$. In particular, fitting to
$1000> \ell > 600$ for the values used in figure \ref{fig:freq} yields $-(d-3)/2$ to machine precision. Typical examples of $\Delta \omega_{j,\,n}$ data and the best fit curve are shown in  the left panel of Fig. \ref{fig:freq} for $d=4$, overtone $n=0$, and horizon radius $r_+=0.1 L$, for the several perturbation sectors we study. On the other hand, in the right panel of this figure, we focus on the scalar gravitational sector of perturbations and study the evolution of the spectrum with the spacetime dimension $d$. A single illustrative example is shown, but we have explicitly checked the power law decay \eqref{wSchAdS} for all perturbation sectors up to $d=11$.
When reading these plots one should bear in mind that the WKB approximation expressions are valid only  for large $\ell_j$, although numerical computation \cite{Berti:2009wx} shows the WKB expression for the scalar field to be a good approximation even for relatively small values of $\ell_S$.

%%%%%%%%%%%%%%%%%%%%%%%%
\subsection{Long-lived quasinormal modes: Rotating black hole \label{sec:WKBmp}}
%%%%%%%%%%%%%%%%%%%%%%%%

It is natural to ask what effects arise from introducing angular momenta, both in the background and in the azimuthal mode number. The most natural background in which to address this question would be the Kerr-AdS black hole or its higher-dimensional generalization, the single spinning Myers-Perry$-$AdS black hole. Unfortunately, an exact harmonic decomposition of general perturbations is not available for these backgrounds. So we instead choose to study the equal angular momenta Myers-Perry$-$AdS (MP-AdS) black hole in odd dimensions.  As a consequence of its enhanced symmetry (it is cohomegeneity-1),
exact analytic results are available both for the angular eigenfunctions and for the associated eigenvalues. This allows much of the WKB quasinormal mode analysis to be done analytically, with numerics playing a very minor role.   The hope is that the key features of their spectrum, such as the deviations of the frequencies from AdS, extend to other rotating AdS black holes.

The equal angular momenta MP-AdS black holes have the following line element  \cite{Hawking:1998kw,Gibbons:2004uw,Gibbons:2004js}
\begin{equation} \label{eq:MPads}
  ds^2 = -f(r)^2dt^2
+g(r)^2dr^2 + h(r)^2[d\psi +A_a dx^a - \Omega(r)dt]^2 + r^2
\hat{g}_{ab} dx^a dx^b
\end{equation}
 where
\begin{eqnarray}\label{eq:MPadsAUX}
 && g(r)^2 = \left[1+\frac{r^2}{L^2} - \frac{2M}{ r^{2N}}\left( 1- \frac{a^2}{L^2} \right) + \frac{2Ma^2}{r^{2N+2}}\right]^{-1},
  \qquad h(r)^2 = r^2\left( 1+ \frac{2Ma^2}{ r^{2N+2}} \right), \nonumber \\
  && \qquad f(r) = \frac{r}{g(r)h(r)} , \qquad \Omega(r) = \frac{2Ma}{ r^{2N} h^2}\,,\qquad N=\frac{d-3}{2}\,,
\end{eqnarray}
and $\hat{g}_{a b}$ is the Fubini-Study metric on $\mathbb{CP}^{N}$ with
Ricci tensor  $\hat{R}_{ab} =2(N+1) \hat{g}_{ab}\,$, and $A = A_a
dx^a\,$ is related to the K\"ahler form $J$ by $dA=2J$.

Asymptotically, the solution approaches the Einstein static universe. The event horizon is located at $r=r_+$ (the largest real root of
$g^{-2}$), which is a Killing horizon generated by the null tangent vector $\partial_t +\Omega_H \partial_\psi$, where the angular velocity is
\begin{equation} \label{angvel}
 \Omega_H = \frac{2Ma}{r_+^{2N+2}+2Ma^2}
 \leq \Omega_H^{\rm ext}\,, \qquad \hbox{where}\quad \Omega_H^{\rm ext}=\frac{1}{L} \sqrt{ 1 + \frac{N L^2}{(N+1) r_+^2} } \,.
\end{equation}
The solution saturating the bound on $\Omega_H$ is an extreme black hole with a regular, but degenerate,
horizon. Note that the upper bound in $\Omega_H\,L$ is always
greater than one and tends to unity in the limit of large
$r_+/L$.

Consider now the Klein-Gordon equation for a massive scalar field \eqref{KGeq}
in this background, which was studied in detail in  \cite{Kunduri:2006qa}.
Assume the separation ansatz for the scalar field,
\begin{equation}\label{MPScalarAnsatz}
\Psi(t,r,\psi,{\bf x})=\sum_{\ell,\,m}e^{-i\,\omega t} e^{i\, m\psi} \,Y_{\ell m}({\bf x})\,  h^{-\frac{1}{2}}\,r^{-N} \, \Phi_{\omega \ell m}(r)\,,
\end{equation}
where $m$ must be an integer since $\psi$ has period $2\pi$, and $Y_{\ell m}({\bf x})$ accounts for the dependence of the perturbation on the $\mathbb{CP}^{N}$ base space coordinates ${\bf x}$. A remarkable (and very useful) property of the equal angular momenta MP black hole is that all the information about the eigenfunctions $Y_{\ell m}({\bf x})$ and its associated eigenvalues are known explicitly \cite{pope:00,Kunduri:2006qa}. This situation is to be contrasted with the Kerr-AdS or single spinning MP cases where the AdS spheroidal harmonics and eigenvalues have no useful analytic expression. More concretely, $Y_{\ell m}({\bf x})$ is an eigenfunction of the charged
scalar Laplacian on $\mathbb{CP}^{N}$ \cite{pope:00,Kunduri:2006qa},
\begin{equation}\label{MPLaplacian}
 -{\cal D}^2 Y_{\ell m}({\bf x}) = \lambda_{\ell m} Y_{\ell m}({\bf x})\,,\qquad \hbox{\rm where} \qquad {\cal D}_a \equiv D_a - i m A_a
\end{equation}
is the gauge-covariant derivative on $\mathbb{CP}^{N}$.  Its eigenvalues are
\cite{pope:00,Kunduri:2006qa}
\begin{equation}\label{eigenCPN}
\lambda_{\ell m} = \ell (\ell+2N) - m^2, \qquad \ell = 2k + |m|, \qquad k=0,1,2,\ldots\,, \qquad |m|\leq \ell.
\end{equation}
With the ansatz \eqref{MPScalarAnsatz}, the Klein-Gordon equation reduces to the angular equation \eqref{MPLaplacian} and to a radial equation for $\Psi_{\omega\ell m}(r)$. The latter can be written in the form of a time-independent Schr\"odinger equation,
\begin{eqnarray}\label{MPKG}
 &&\hspace{-1cm} \partial_z^2\Phi_{\omega\ell m}+\left(\omega-U_{+} \right)\left(\omega-U_{-} \right)\Phi_{\omega\ell m}=0\,,\qquad
 V_0=\frac{f^2\sqrt{h}}{r^{N+1}}\frac{d}{dr}\left[\frac{f^2 h}{r}\frac{d}{dr}\left(\sqrt{h}r^{N}\right) \right],
\\
 &&U_{\pm }=m \Omega \pm \sqrt{
 V_0
+\frac{f^2}{r^2 h^2 } {\biggl \{}   m^2 r^2+h^2\left[\ell(\ell+2N)-m^2+r^2 \mu ^2+4 (1-\sigma ) \left(\frac{h^2}{r^2}-1\right)\right]   {\biggr \}}}, \nonumber
\end{eqnarray}
where $\sigma =1$ (for the scalar field; see below), and  we have introduced the tortoise coordinate
\begin{equation}\label{MPTartoise}
z=\int_r^\infty dr\,\frac{g}{f}\,, \qquad 0\leq z< +\infty
\end{equation}
that describes the horizon at $z\to +\infty$ and the asymptotic boundary at $z\sim L^2/r \to 0$.

So far we have discussed a massive scalar field in the background of our rotating black hole. We would also like to consider perturbations of the background gravitational field. In order to disentangle perturbations of the Schwarzschild black hole, KI classified all perturbations according to how they transformed under isometries of the $(d-2)$-sphere. Similarly, gravitational perturbations of the MP-AdS black hole \eqref{eq:MPads}  can be decomposed into scalar, vector and tensor parts, according to how they transform under isometries of the $\mathbb{CP}^N$ base space. The details of these perturbations are discussed in \cite{Kunduri:2006qa}. The relevant information for our own study is that, for tensor gravitational perturbations, solving  the linearized Einstein equations reduces  to solving a radial equation for a single gauge invariant quantity. This equation can be written exactly in the Schr\"odinger form \eqref{MPKG} for $\mu=0$ (and $\sigma\neq 0$) \cite{Kunduri:2006qa}.
Tensor perturbations exist only for odd $d\geq 7$, \ie integer $N\geq 2$.

To summarize, the radial Schr\"odinger equation \eqref{MPKG} describes perturbations of a massive scalar field (for $\sigma=1$ and $d\geq 5$) and, in addition, tensor gravitational perturbations (for $\mu=0$ and $d\geq 7$).\footnote{We refer the reader to \cite{Kunduri:2006qa} for a discussion of further constraints on $\sigma$ and on the eigenvalues that are irrelevant here.}

We are interested in the large angular momentum WKB limit, $\ell \to\infty$, of the radial Schr\"odinger equation \eqref{MPKG} .
An important difference with respect to the non-rotating case discussed previously is that the behavior now depends on the azimuthal quantum number $m$.   The most interesting WKB modes have $|m|\sim \ell \to\infty$ which we henceforward assume to be the case. Following closely the non-rotating computation of the previous section, we define
\begin{equation}\label{SwkbMP}
\Phi_{\omega\ell m}(z)=e^{p\, S_{w p}(z)} \,,\quad  \omega=p\,w\,,   \quad \ell=p -N\,,\quad m=\epsilon_\psi \ell \;\;(\epsilon_\psi=\pm 1)\,,\quad \nu_j=\sqrt{(N+1)^2+\mu_j^2L^2}\, ,
\end{equation}
where $\mu_j=\mu$ for the massive scalar field and $\mu_j=0$ for tensor gravitational perturbations.
Equation \eqref{MPKG} then reads
\begin{eqnarray}\label{SchroEqWKBmp}
&&\hspace{-1cm}\frac{1}{p} S_{w p}''(z)+S_{w p}'(z)^2-\left[V_+(z)-w\right] \left[V_-(z)-w\right]=\frac{1}{p^2}\,\tilde{\chi}_j(z)\,,\quad \hbox{with} \quad  V_{\pm }(z)=\epsilon_\psi \Omega\pm\frac{r}{g h^2}\,, \nonumber \\
&&
 \end{eqnarray}
and $\tilde{\chi}_j(z)$ is a function associated to the higher order WKB contribution whose explicit expression can be read from \eqref{MPKG} but is irrelevant for our study. Notice that $V_{\pm}\to \sqrt{V}$ defined in \eqref{SchroEqWKB} when  $\Omega_H \to 0$.
Note also that the harmonic decomposition of this subsection is done with respect to the $\mathbb{CP}^N$ base space. When sending $\Omega_H \to 0$ and comparing results, we should have in mind that the $\mathbb{CP}^N$  harmonics form a subset of the KI ($S^N$) harmonics.

\begin{figure}[t]
\centerline{\includegraphics[width=.50\textwidth]{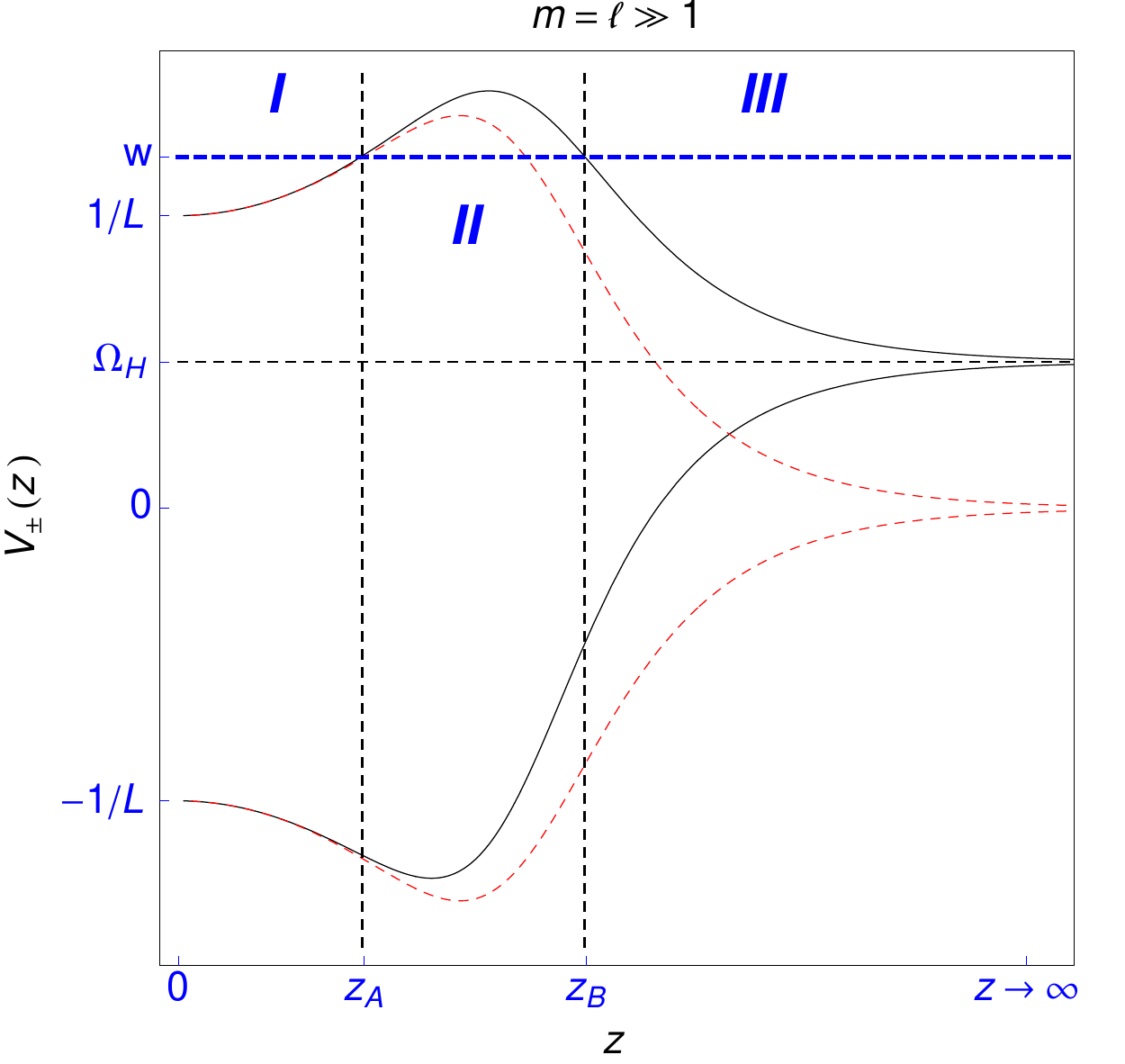}
\hspace{0.5cm}\includegraphics[width=.50\textwidth]{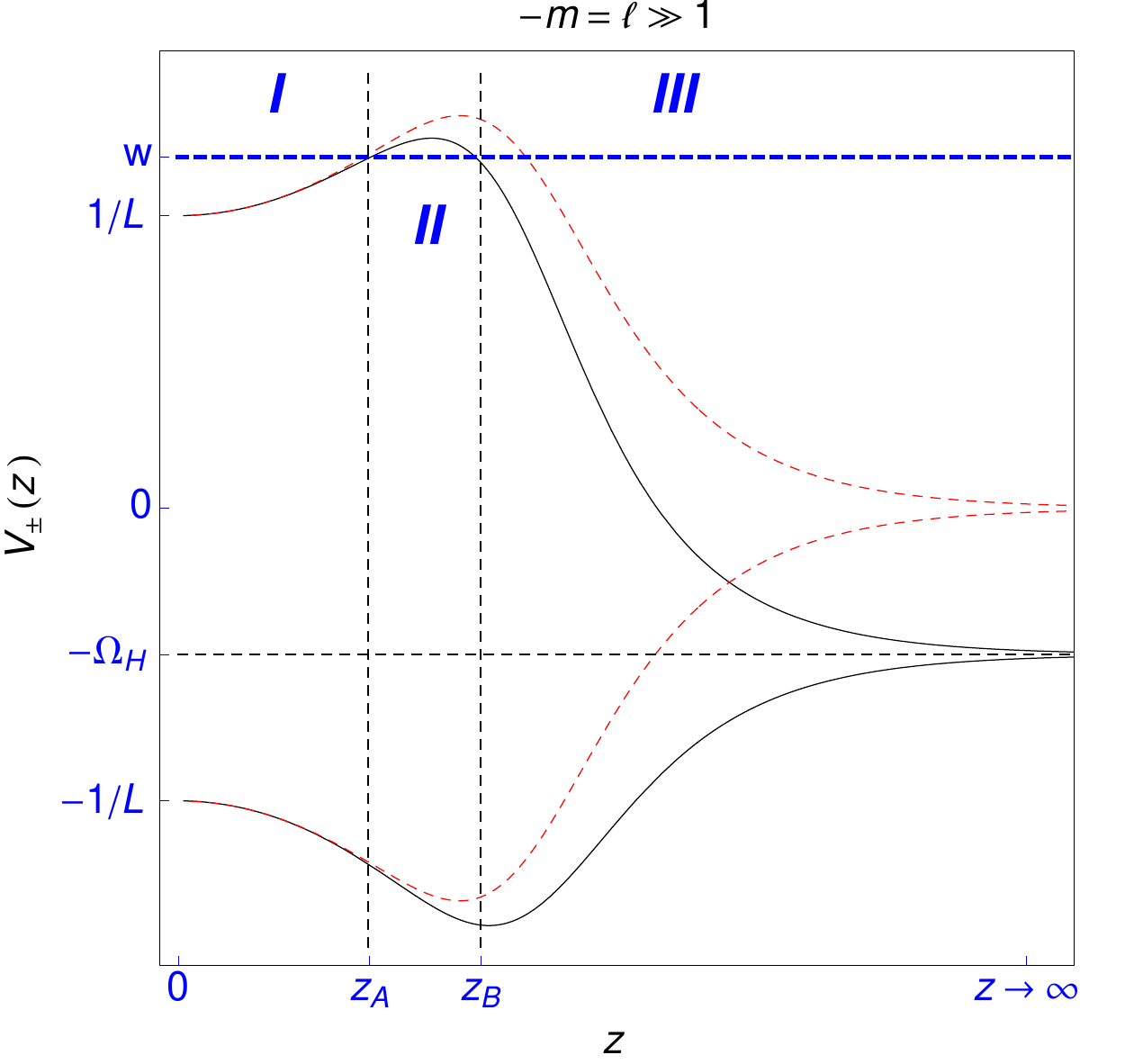}}
\caption{The black solid line describes the WKB potentials $V_{\pm}(z)$ for the rotating MP black hole (in the non-superradiant regime,  $\Omega_H\, L <1$).
For comparison, the red dashed lines describe the same potential when the rotation vanishes, \ie the Schwarzschild-AdS limit. We also show the  WKB allowed ($I,III$) and forbidden ($II$) regions for a given (rescaled) frequency $w$. The left panel describes corotating modes ($m=\ell \gg 1$), while the right panel is for counter-rotating modes ($-m=\ell\gg 1$)} \label{fig:MPwkb}
\end{figure}

The quasinormal mode frequencies in the limit of large $p$ can be found by solving \eqref{SchroEqWKBmp}   in a WKB approximation. The procedure is now very similar to the Schwarzschild-AdS case analyzed in the previous section. The form of the potentials $V_\pm$ is illustrated in  Figure \ref{fig:MPwkb}.
It is well-known that black holes with $\Omega_H\,L >1$ are superradiantly unstable (\ie its ``quasinormal" modes grow in time in this regime). Here, we are interested in the case where this instability is absent,  $\Omega_H\,L <1$; this is the case addressed in Figure \ref{fig:MPwkb}.
 Under this condition, long-lived quasinormal modes do exist. They are bound states in the exterior allowed region $I$ that can decay through the potential barrier (region $II$) downwards toward the black hole horizon located in the WKB region $III$. In the left panel plot, the upper black solid line describes the WKB potential $V_{+}(z)$ when the wave co-rotates with the black hole, \ie when $m=\ell\gg 1$. We compare it with the same potential in the limit $\Omega \to 0$, where we recover the Schwarzschild-AdS case of the previous subsection. This is the red dashed line. In the rotating geometry, the co-rotating modes see a potential barrier that is both higher and wider than the non-rotating one (for a given frequency $w$ and $\ell$). Therefore, we anticipate that in the rotating background corotating quasinormal modes can be even more long lived than in  Schwarzschild-AdS. The plot for $V_+(z)$ in the right panel shows that the opposite conclusion applies to counter-rotating modes that have  $-m=\ell\gg 1$. To confirm and quantify these expectations we do the actual WKB analysis. Fortunately, we can borrow most of our calculations of the previous subsection, namely equations \eqref{wkb1z0} and \eqref{wkb1zfinite}-\eqref{QNcondition}, as long as we take the replacement
$Q_j(z)\equiv w_j^2 -V(z) \to \left[w_j-V_+(z)\right] \left[w_j-V_-(z)\right]$, and $\nu_j$ as defined in \eqref{SwkbMP}.

In the end, we find the real part of the frequencies of the quasinormal mode spectrum are quantized by the condition
\begin{equation}
\eta (w_j) \equiv n\,\pi +\frac{\pi}{4} \quad  \Leftrightarrow \quad  p \int_0^{z_A} \! \! \! \! \sqrt{\left(w_{j,\,n}-V_+\right)\left(w_{j,\,n}-V_-\right)}\, dx  =\frac{\pi}{2} (2n+1+\nu_j)\,, \quad n=0,1,2,\cdots
 \label{ressonance freqMP}
\end{equation}
with $\nu_j$ defined in \eqref{SwkbMP} ($\mu_j=\mu$ for the massive scalar field and $\mu_j=0$ for the tensor gravitational perturbation).
The damping timescale scales as
\begin{equation}
\tau_{j,\,n} \sim \exp \left( 2 p \int_{z_A}^{z_B}\!\! \sqrt{\left(V_+ -w_{j,\,n}\right)\left(V_- -w_{j,\,n} \right)}\,dx\right), \quad n=0,1,2,\cdots
 \label{damptimeMP}
\end{equation}

It follows that at large angular momentum the modes are again long-lived with a damping timescale \eqref{damptimeMP}  that scales as $\tau\sim e^{\ell}$.
Furthermore, evaluating \eqref{ressonance freqMP} numerically and fitting $\Delta \omega_{j,n}$ to a power law again yields \eqref{wSchAdS}, though with constants $K_d^{(j)}$ which now depend on the angular momentum of the black hole background.  We have checked this for $d = 5,7,9,11$ and for both co-rotating and counter-rotating modes.  This supports the conjecture that this behavior (with appropriately solution-dependent $K_d^{(j)}$) is universal for solutions asymptotic to global AdS.

%%%%%%%%%%%%%%%%%%%%%%%%%%%%%%%%%%%%%
\section{Nonlinear stability of geons and boson stars}
\label{approx}

We now investigate the stability of horizon-free stationary solutions such as geons and boson stars which are asymptotic to global AdS solutions.   The modes most likely to drive such an instability are those with large angular momentum $\ell$ and small radial excitation, since they have frequencies close to modes of AdS. Indeed,
in the previous section we saw that for AdS${}_{d}$ Schwarzschild, such modes have frequencies (\ref{wAdS}, \ref{wSchAdS})
%  \footnote{The notation below is adapted to the case with time-reversal symmetry, though the behavior for Kerr is the same.}
\be
\label{Kfreq}
L\,\omega = L \omega_{AdS} + {\cal O}(\ell^{-\frac{d-3}{2}}) = \ell + C + {\cal O}(\ell^{-\frac{d-3}{2}})
\ee
where $C$ is a constant that depends on dimension and the type of perturbation.   By examining a class of higher dimensional rotating black holes, we  saw that rotation does not change the last term. This suggests that we will find similar behavior for perturbations about generic stationary solutions asymptotic to global AdS (see again footnote \ref{sources}).  In particular, horizon-free stationary solutions should exhibit good approximations to the resonances that drive the nonlinear instability in AdS.

Whether or not there is an instability will depend on how much power is initially contained in these large $\ell$ modes. This is directly related to the degree of differentiability of the perturbations we choose to consider, and can be quantified as follows.
Suppose we have a field on ${\mathbb R}^{d-1}$  with Fourier coefficients $\tilde \phi (k) \sim k^{-p}$. To be in the Sobolev space $H^s$ we need
\be\label{sobelev}
\int d^{d-1} k [k^s\tilde \phi (k)]^2 < \infty.
\ee
This requires
\begin{equation}
\label{pconstraint}
p > s + \frac{d-1}{2}.
\end{equation}
 We certainly want  to choose $s$ so that our perturbations would induce no instability of Minkowski spacetime. In Einstein-Hilbert gravity, four dimensional Minkowski spacetime has been shown \cite{Bieri:2009xc} to be stable for $s=3$ and is further believed \cite{private} to be stable even for $s>5/2$. In $d$ dimensions, it is believed \cite{private} to be stable for $s>(d+1)/2$. From (\ref{pconstraint}), this requires $p>d$.
By the Sobolev embedding theorem, a field is in $C^q$ (\ie $q$ times differentiable) if $p> q + d-1$. So the minimum requirement $p>d$ ensures that the fields are in $C^1$ but not necessarily $C^2$.
We will find that geons and boson stars are nonlinearly stable unless they are in a high dimension and perturbations have low differentiability.

We wish to add a small perturbation to a given solution and study its time evolution.  We will address this problem by attempting to construct the general nonlinear solution using formal classical perturbation theory. By formal, we mean that we will focus on the existence of a good asymptotic series rather than attempting to demonstrate convergence.  We will see that this level of argument predicts results that agree qualitatively with rigorous theorems for the nonlinear Schr\"odinger equation as reviewed in section \ref{NLSE}.

Recall that the perturbative expansion can be represented as a sum over connected tree graphs, where the vertices of the graph are associated with the monomials that appear in a Taylor series expansion of the action about the given background and the lines of the graph are associated with a given propagator. As usual, it is simplest to work in terms of normal modes. We  assume that we are given a ``seed'' solution of the linearized equations of motion written as a sum over normal modes $\psi_{\vec \ell}$, where $\vec \ell$ represents a complete set of labels for the normal modes: $\phi^{(1)}(x,t) = \sum_{\vec \ell} \phi^{(1)}_{\vec \ell} \psi_{\vec \ell}$.
We seek a solution of the form
\begin{equation}
\phi (x,t) = \sum_{\vec \ell} \phi_{\vec \ell}(t) \psi_{\vec \ell}(x),
\end{equation}
with\footnote{The $n$ in this section should not be confused with the one in the previous section. Here $n$ will represent the order of the perturbation expansion and not the radial excitation of the linearized mode.}
\begin{equation}\label{perttheory}
\phi_{\vec \ell}(t) = \sum_{n=1}^\infty \phi^{(n)}_{\vec \ell}(t).
\end{equation}
Here each $\phi^{(n)}_{\vec \ell}$ is an order $n$ homogeneous polynomial in the first order  coefficients $\phi^{(1)}_{\vec \ell {}'}$ and
at each order $n$ we require the equations to be satisfied only up to higher order terms.

For $n>0,$
the terms of the polynomial $\phi^{(n)}_{\vec \ell}$ are associated with tree graphs with $n+1$ external lines as follows. For each such graph, label one external line with the pair $(\vec \ell,+)$.  This line is to be thought of as the root of the tree (taken to be at the bottom) and is associated with the mode we are trying to compute.  The other external lines are assigned arbitrary pairs $(\vec \ell_j, \pm_j)$ for $j = 1, \dots, n$ of modes and signs\footnote{The inclusion of both signs stems from the fact that our modes are real and thus contain both positive and negative frequencies.}.  We may think of these as the leaves of the tree (and we take them to be at the top).   Internal lines are also assigned arbitrary pairs of modes and signs.  Each vertex  $v$ contributes a coupling constant $g$, an overlap factor $C(\{ \vec \ell \})$, and an energy denominator
\be
\frac{1}{\sum_{{\rm lines} \ i \in v} \tilde  \omega_i},
\ee where we say that $i \in v$ for a line $i$ and vertex $v$ if $i$ is attached to $v$.    The sum in the energy denominator has one term for each line connected to the vertex.  For an external line or for the internal line below the designated vertex we set $\tilde \omega_i = \pm_i \omega_i$ where $\pm_i,$ $\omega_i$ are the signs and frequency associated with the mode assigned to that line. For an internal line above the given vertex $v$, we define $\tilde \omega_i$ to be the (signed) sum $\sum_{{\rm leaves \ }j > i} \pm_j \omega_j$ over all external lines $j$ that lie above $i$ in the tree ($j > i$).  In a theory without derivative couplings the overlap factor $C(\{ \vec \ell \})$ is just an integral over space of the product of all modes associated with the lines connected to the given vertex, though in a gravitational theory two of these modes will appear with derivatives.

These factors are multiplied together, and also multiplied by $\phi^{(1)}_{\vec \ell_1} \dots \phi^{(1)}_{\vec \ell_n} \exp(-i \sum_{j=1}^n  \tilde \omega_j t).$
The result is then summed over all non-trivial\footnote{A set is trivial if the corresponding sum vanishes no matter what frequencies are assigned to each mode.  E.g. $\omega_{\vec \ell_i} - \omega_{\vec \ell_i} + \omega_{\vec \ell_j} - \omega_{\vec \ell_j} $ at third order.  Excluding these terms is associated with a resummation of naive perturbation theory which shifts the frequencies of each mode at each order but which should not affect the asymptotics \eqref{Kfreq} in the limit of small perturbations.} sets of pairs of signs and modes.  For example, the second order term in any theory comes just from the cubic vertex and takes the form
\begin{equation}
\phi^{(2)}_{\vec \ell} = g \sum_{\pm_1, \vec \ell_1, \pm_2, \vec \ell_2} \phi^{(1)}_{\vec \ell_1}  \phi^{(1)}_{\vec \ell_2} C(\vec \ell_1, \vec \ell_2) \frac{\exp(-i[\tilde \omega_1 +\tilde \omega_2]t)}{\omega +\tilde \omega_1 +\tilde \omega_2}.
\end{equation}

We now wish to estimate the result of the  sum (\ref{perttheory}).  As we will argue below  that for $d \ge 7$ the sum is dominated by the modes with no radial excitation, it is convenient to restrict to both $d \ge 7$ and no radial excitations from the beginning.  At the end of our analysis we will also explain the result for $6 \ge d \ge3$.

As mentioned above, we will focus on whether the sum associated with a given graph is finite; i.e. whether it is UV-convergent.  Since we are interested in the high-momentum behavior we may use the flat-space approximation in which the sums over modes are replaced by integrals over Fourier space.  This has the advantage that the overlap functions $C(\{\vec \ell \})$ become delta-functions that enforce conservation of spatial momentum, perhaps multiplied by certain powers of momentum associated with the derivatives in the given interaction vertex.  Since all diagrams are tree graphs, these delta-functions can be used to eliminate the sum over internal momenta as well as one of the sums over external momenta.  Since we study the sum with only modes on the sphere, each remaining momentum lives in ${\mathbb R}^{d-2}$.

It is useful to organize the momentum integrals in terms of an overall momentum scale $k$ (defined, e.g., by $k^2 = \sum_{i=1}^n |\vec k_i|^2/n$) and a remaining set of integration variables $\{ \kappa_j \}$ which take values in $S^{(d-2)(n-1)-1}$. The contribution of a given diagram of order $n$ in the linear seeds $\phi_{\vec \ell}^{(1)}$ can now be written

\begin{equation}
\label{kint}
{\rm graph}_{n}  \sim \sum_{signs} \int k^{(d-2)(n-1)-1} dk \int_{S^{(d-2)(n-1)-1}} d\kappa \left (\prod_{j=1}^{n} \phi^{(1)}_{\vec \ell_j} e^{-i\tilde\omega_j t }\right) \prod_{v \in {\rm vertices}}  \frac{g_v F_v(k, \kappa)}{\sum_{{\rm lines \ i \in v}} \tilde\omega_i},
  \end{equation}
where $i \in v$ for a line $i$ and vertex $v$ means that $i$ is attached to $v$.

\begin{figure}[!t]
\label{3to4}
\begin{center}
\vspace {-5pt}
\includegraphics[scale=0.5]{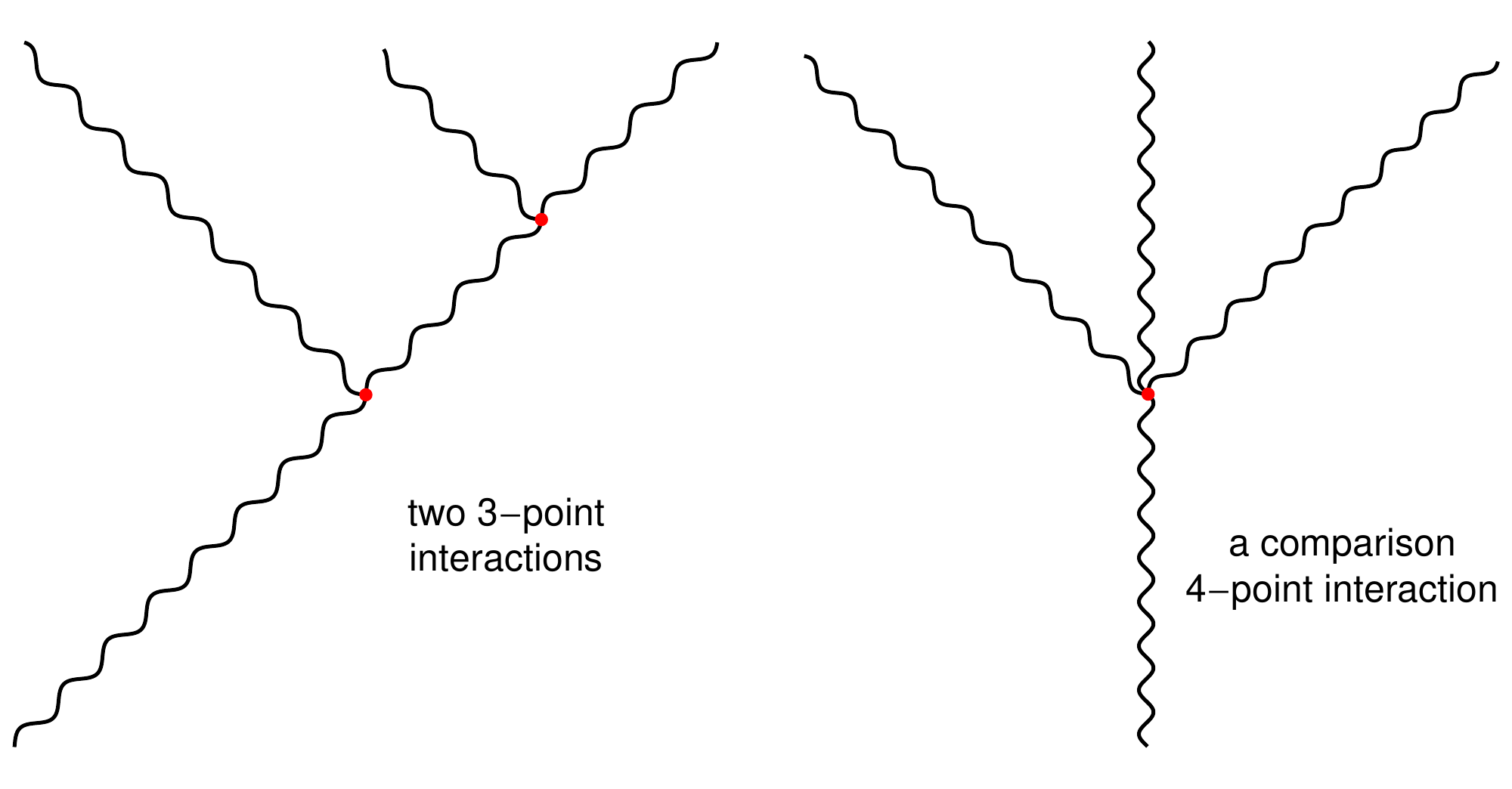}
\end{center}
\vspace {-10pt}
\caption{\label{fig:3to4}Two 3-point interactions combine to make an effective 4-point vertex.
}
\end{figure}

For Einstein-Hilbert gravity coupled to a scalar field, the largest interactions in the UV  have two derivatives. So we may write $F_v(k, \kappa) = k^2 \tilde F_v(\kappa)$.  We may also write $\phi^{(1)}_{\vec \ell_i} =  k^{-p} \tilde \phi_i(\kappa)$.   Now, the gravitational action contains vertices of arbitrary order.  However, as noted above, each vertex is associated with precisely two derivatives.  On the other hand, we may use the cubic vertex to build an effective higher-order vertex of any order;  figure \ref{fig:3to4} shows a 4-point example.  Since the effective $m$-point vertex is built from $m-2$ cubic vertices, it includes a factor of $k^{2(m-2)}$ from derivative couplings.   While it also contains $m-3$ extra energy denominators, each denominator is no larger than ${\cal O}(k)$.  So the composite $m$-point vertex built from cubic vertices is at least ${\cal O}(k^{m-3})$ times the size of the fundamental $m$-point vertex at large $k$.  Thus we see that the dominant graphs at large $k$ involve only cubic vertices.   Note that at order $n$ in the linear seeds $\phi^{(1)}_{\vec \ell}$ such graphs involve $n-1$ cubic vertices\footnote{At order $n$ we have $n+1$ external lines. Since each internal line connects to two vertices, connected tri-valent graphs satisfy $E + 2I = 3V$, where $V$ is the number of vertices, $E$ is the number of external lines, and $I$ is the number of internal lines.  But for any tree graph $V = I+1$ by Euler's theorem.
Thus $I=n-2$ and $V=n-1$.}.   We therefore neglect all other graphs for the rest of this section and write

\begin{equation}
\label{kint2}
{\rm graph}_{n}  \sim \sum_{signs} \int k^{d(n-1)-1 -np}   dk \int_{S^{(d-2)(n-1)-1}} d\kappa \left (\prod_{j=1}^{n} \tilde \phi_j(\kappa)  e^{-i\tilde \omega_j t }\right) \prod_{v \in {\rm vertices}}  \frac{g_v \tilde F_v(\kappa)}{\sum_{{\rm lines \ i \in v}} \tilde\omega_i}.
  \end{equation}

It remains to estimate the energy denominators.  From \eqref{Kfreq}, we see that at large $\ell$ each $\omega_i = {\cal O}(k)$.  A typical energy denominator thus also satisfies ${\sum_{{\rm lines \ i \in v}} \tilde \omega_i} \sim {\cal O}(k)$.
However, in some cases the ${\cal O}(k)$ terms can cancel in the energy denominator.  From \eqref{Kfreq} we see that this happens precisely when the sum
${\sum_{{\rm lines \ i \in v}} \tilde \omega_i}$ vanishes in the approximation $\omega_i = \ell_i$, where $\ell_i$ is the integer associated with the total angular momentum of the mode $i$.  The condition that this occurs may be used to express e.g. $\ell_2$ in terms of the other angular momenta, reducing the integral over $\kappa$ to $(d-2)(n-1)-2$ dimensions, and similarly for each additional constraint imposed.  We note that exact resonances of this form were demonstrated in \cite{Bizon:2011gg,Dias:2011ss} for perturbations about empty AdS.

An important observation is that, when this condition is imposed, the ${\cal O}(1)$ term in \eqref{Kfreq} cancels whenever (when written directly in terms of external momenta!) the set of signs $\{\pm_i\}$ contains equal numbers of positive and negative signs. Note that making this choice of signs imposes no further constraints on the momenta.  So, by making this choice of signs and imposing the above-mentioned single constraint we now find that ${\sum_{{\rm lines \ i \in v}} \tilde\omega_i} = {\cal O}(k^{-(d-3)/2})$ or smaller.  For this reason, momenta satisfying this constraint should dominate the result: While imposing the constraint reduces the integral by one dimension, this will ``cost" no more than a factor of roughly $k$.  On the other hand, we have increased the contribution from the energy denominator by $k^{(d-1)/2}$. This is precisely the effect of the approximate resonances.

Note, however, that such a choice of signs can be made only for energy denominators that involve an even number of external momenta.  One might think that, when an odd number of external lines are involved, one could simply shift the constraint to require e.g. $\sum_i (\pm_i) \ell_i = C$ instead of $\sum_i (\pm_i) \ell_i = 0$.  However, the
situation is not so simple.  Indeed, refs. \cite{Bizon:2011gg,Dias:2011ss} studied the explicit form of perturbation theory about empty global AdS${}_4$ and found no resonances of this form at second order (three external lines).  In simple cases (e.g. for three scalar-type gravitational modes) elementary considerations of $SO(d-2)$ representation theory force the associated overlap functions to vanish, removing any potential resonances.  While we have not been able to rule out all potential 2nd order resonances in this way, we have explicitly studied all analogous cases involving a graviton (of any type) and two modes of a massive scalar field (of any mass) and found no resonances at this order.

It is likely that new approximate resonances of the above form do in fact arise at higher even orders in perturbation theory.  But these new resonances turn out to have little effect.  We now make a short digression to explain why. The point is that, for a given diagram of any order, no more than half of the energy denominators can be approximately resonant in the above sense.  This can be seen from the fact that our perturbation theory is dominated by diagrams built entirely from 3-point vertices as follows.  

First recall that, in computing the relevant $\tilde \omega_i$ that contributes to an energy denominator, the definition of $\tilde \omega_i$ for a given internal line depends on whether this line lies above or below the vertex in question.  However, for a given labeling of the external lines, the two definitions coincide
precisely when vertex above the line has an exact resonance.  When this coincidence occurs, we refer to the line in question as a resonant line.  We may similarly speak of approximately resonant internal lines using the above sense of approximate resonances.  Now, imagine dividing a given diagram into subdiagrams by cutting all of the approximately resonant internal lines.  The subdiagram containing the original root merits special treatment: if the root (an external line) is non-resonant we will discard this subdiagram, though we keep it if the root is resonant.
 
With this understanding the (approximately) resonant energy denominators for the original diagram are in one-to-one correspondence with the subdiagrams obtained above,  so that counting one also counts the other. Due to the above coincidence regarding $\tilde \omega_i$, each subdiagram comes with a definite labeling of its external lines by (perhaps approximate) modes.  Since by construction each such subdiagram contains a resonance, it must contain more than one 3-point vertex.  That is, the number of sub-diagrams is no more than half of the number of vertices. This verifies the claim that the number of approximate resonances is also bounded by half the number of vertices, i.e.,  there are at most $n$ of the above approximate resonances at either order $2n+1$ (with $2n$ vertices) or $2n+2$ ($2n+1$) vertices.

Returning to the main calculation, the above argument motivates  what we call the generic frequency assumption.  Namely, we will assume that the integral \eqref{kint} is dominated\footnote{Or, at least `marginally dominated' in the sense that the contributions from such configurations are of the same order at large $k$ as the full answer.  That is, we may neglect other sets of configurations which give contributions of the same order, such as those obtained by not imposing the constraint $\sum_i (\pm_i) \ell_i=0$ for an energy denominator with an odd number of external momenta.  When all phases are aligned, this would allow the phase space volume to contribute an additional factor of $k$  and would reduce the contribution of the energy denominator by only a compensating factor of $k$. } by configurations in which, for each of the ($n-1$) vertices, the constraint $\sum_i (\pm)_i \ell_i =0$ is satisfied.

We furthermore assume that the integral is dominated by configurations (of momenta and signs) for which the contribution due to each energy denominator with an even number of terms is ${\cal O}(k^{(d-3)/2})$ and for which the contribution due to each energy denominator with an odd number of terms is ${\cal O}(1)$ at large $k$.   We then write $\sum_{{\rm lines \ i \in v}} \tilde\omega_i = S_v(\kappa), k^{-(d-3)/2}S_v(\kappa)$ as appropriate.  As noted above, there are at most $n$ approximately resonant vertices at either order $2n+1$ or $2n+2$.  We therefore focus on odd orders ($2n+1$) which will clearly show the strongest UV effects.

Letting $\tilde \kappa$ denote configurations that satisfy the constraints associated with the above-mentioned generic frequency assumption, we may now estimate \eqref{kint2} at order $2n+1$ to be
\begin{equation}
\label{kint3}
{\rm graph}_{2n+1} \sim \int  k^{n(d-3)/2} k^{2nd -(2n+1)(p+1)}   dk \int_{\Sigma^{(d-3)(2n-1)-1}} d\tilde{\kappa} \left (\prod_j^{2n} \tilde \phi_j (\tilde{\kappa})  e^{-i\tilde\omega_j t }\right) \prod_{v \in {\rm vertices}}  \frac{g_v \tilde F_v(\tilde{\kappa})}{S_v(\tilde{\kappa})},
\end{equation}
where $\Sigma^{(d-3)2n-1}$ is a ${(d-3)2n-1}$ dimensional manifold defined by imposing the required constraints on $S^{(d-2)(2n-1)-1}$.
A similar result holds at order $2n$.

We are now in a position to justify our restriction to modes with no radial excitation for $d \ge 7$.  A study of black hole quasinormal frequencies for modes with radial excitations indicates that they admit an expansion $\Delta \omega = \ell + C_0 + C_1/\ell + \dots$ at large $\ell$ with coefficients $C_i$ that depend on the radial quantum number.  Let us consider the effect of radial excitations at third order in perturbation theory, where we have one energy denominator that refers to 4 modes.  As usual, we take two to be of each sign to enhance cancellations.  However, the constant term $C_0$ cancels only when the radial quantum numbers agree pairwise.  That is, the two `positive' modes have the same radial quantum numbers as the two `negative' modes.    This imposes two constraints on the 3 sums over radial quantum number.   So the extra phase space associated with the sum over radial modes with radial quantum number $\ll   \ell$ can contribute no more than a factor of $\ell$. Furthermore, we generally find an energy denominator of order $1/\ell$ as opposed to the $1/\ell^{\frac{d-3}{2}}$ obtained in the case with no radial excitations.  For $d \ge 7$, we see that the sum is indeed (at least marginally) dominated by the restricted sum over modes with no radial excitations.  The same effect occurs at higher orders.

The last ingredient to consider are the phases $e^{-i\tilde\omega_j t} $ and the phases in $\tilde \phi_i(\kappa)$.    The result is largest when all of the oscillatory terms are in phase; e.g. when the $\tilde \phi_j e^{-i\tilde\omega_j t} $ are all real and positive.  This may take an exponentially long time to achieve, but since we are interested in late time stability, we will estimate the result assuming this condition is satisfied.

In this case the $\tilde\kappa$ integral becomes independent of $k$ and we find

\begin{equation}
\label{kint4}
{\rm graph}_{2n+1} \sim  \int k^{n(d-3)/2}  k^{2nd -(2n+1)(p+1)} dk .
\end{equation}
The integral over $k$ converges in the UV if
\begin{equation}
(d-3)/2 + 2(d-1) -2p  < p/n .
\end{equation}
 So for
 \be
 \label{pres}
 p > \frac{5d-7}{4}
 \ee
it converges for all $n$ while it diverges at large $n$ for smaller $p$.  The terms at order $2n$ are slightly more convergent at large $k$, so the UV properties are determined by the odd terms \eqref{kint4}. Note that when the integral converges in the UV, terms of higher order in $n$ scale with higher powers of the seed solution's amplitude and we have a good asymptotic series. In such cases it then appears that the solutions are nonlinearly stable over arbitrarily long time periods for initial data that lies in a Sobolev space $H^s$ for $s> (3d-5)/4$ (using (\ref{pconstraint})). The similarity of this conclusion to theorems regarding the nonlinear Schr\"odinger equation (reviewed in section \ref{NLSE}) encourages the belief that it can also established more rigorously.

Recall that to avoid an instability in Minkowski spacetime, we need at least $p > d$.
Our analysis above indicates a divergence with $p > d$ when $(5d-7)/4  > d$ or $d > 7$, \ie starting at AdS${}_8$.   But as discussed at the beginning of this section, even that requires perturbations which are only $C^1$ and not $C^2$. To see an instability with perturbations that are $C^2$, one needs to go to at least to AdS${}_{12}$.  One may in fact need to go even higher.  As noted in \cite{Dias:2011ss}, the detailed structure of the gravitational interactions can cause certain overlap coefficients $C(\{ \vec \ell \})$ to vanish, removing potential (in this case approximate) resonances and making the sum more convergent than our analysis would indicate.   It is also possible that further resummation of the series (analogous in the case of exact resonances to introducing additional frequency shifts at each order in perturbation theory) might further improve convergence.  One should thus take our results as giving a lower bound on the dimension required for instability at a given degree of differentiability.

The analysis above was valid only for $d \ge 7$.  For smaller dimensions one must include the sum over radial quantum numbers.  As noted above, this contributes an extra factor of $k^{\frac{7-d}{2}}$ at third order.  For each additional two orders, it contributes an extra factor of $k$ from the additional phase space, but reduces the integrand by a factor of $k^{\frac{d-5}{2}}$ due to the larger energy denominator,  i.e., at order $2n+1$ it contributes $k^{\frac{n(7-d)}{2}}$. The analog of \eqref{pres} is then $p > d$, which is the same condition under which Minkowski space is expected to be unstable.

Even in cases of high dimension and low differentiability, one needs to wait for the phases to line up to see an instability. A more detailed estimate of (\ref{kint3}) which includes random phases shows that the integral over $k$ always converges in all dimensions assuming only the minimum differentiability required for stability of Minkowski spacetime.

 For AdS black holes, since  the large $\ell$ modes are long lived, we can ask if the above phases are indeed likely to line up before the relevant modes decay.  Since the probability of $N$ phases lining up to accuracy $\epsilon$ is of order $\epsilon^N$, we will generically  need to wait a time of order $\epsilon^{-N}$ before the phases line up.  Since there are $k^{2n(d-3)}$ phases in \eqref{kint3} at momentum scale $k$, these phases will line up only at $t_{\rm in \ phase} \sim \epsilon^{-k^{2n(d-3)}}$.  But on the other hand we have seen that in the presence of a black hole the decay time for modes of momentum $k$ is only $e^k \ll t_{\rm in \ phase}$. So for $d \ge 4$ the phases are not in fact likely to line up before the relevant modes decay.

\section{Non-coalescing binaries}

In this section we describe a new class of asymptotically AdS solutions consisting of non-coalescing black hole binaries. It is easy to see that such solutions should exist. For simplicity, we consider four dimensional spacetimes. Consider a non-extreme Kerr AdS black hole with angular velocity $\Omega > 1/L$. The Killing field $\xi = \partial/\partial t + \Omega \partial/\partial \varphi$ which is null at the horizon is timelike just outside the horizon and spacelike near infinity. So there must be a radius at which $\lambda = \xi_\mu \xi^\mu$, evaluated on the equatorial plane, has  an extremum. The $\theta \rightarrow \pi - \theta$ symmetry ensures that at this radius, $\nabla_\mu \lambda = 0$. Now
\be
\xi^\mu \nabla_\mu \xi^ \nu = -\frac{1}{2} \nabla^\nu \lambda = 0
\ee
at this radius, so the Killing field is actually tangent to a geodesic. This geodesic  represents a circular orbit around the central black hole. Imagine placing a small black hole on this circular orbit. This will perturb the background geometry, but since the orbit of the small black hole is invariant under the Killing field $\xi$, the perturbation generated will also be invariant under this Killing field. Since $\xi$ is the horizon generator of the original Kerr AdS black hole, this means that no energy falls into the black hole at the center. The radiation produced by the small black hole forms a standing wave which just supports the orbit. We have constructed this radiation explicitly and it will be presented elsewhere \cite{Jorge}.  It should be possible to add higher order corrections to this perturbative argument and find exact solutions of this type.

One can ask whether the above circular orbit is stable to radial perturbations. It turns out that if the black hole is too close to extremality, the orbit is unstable, but there is a range of parameters for Kerr AdS so that the orbit is stable. Unfortunately, even when the orbit is stable, the entire solution is still unstable due to superradiance. This is because we need $\Omega > 1/L$ to have a circular orbit which is invariant under the horizon Killing field. Generic perturbations will probably cause the central black hole to spin down and the orbiting black hole will spiral in as usual.  

One might also ask if one can place a small black hole in permanent orbit around a geon, where the radiation produced would again support the orbit against spiraling in. The answer appears to be no.   The Killing field of the geon, $\xi = \partial/\partial t + (\omega/m)\partial/\partial \varphi$ is timelike at the origin and spacelike at infinity. But since there is no analog of the horizon where the Killing field is null, $\lambda = \xi_\mu \xi^\mu$ need not have any extrema\footnote{A continuity argument shows that there are no extrema for geons with small energy.}.

It is amusing to note that for the non-extreme Kerr metric (in asymptotically flat spacetime) the horizon generator $\xi = \partial/\partial t + \Omega \partial/\partial \varphi$ is always timelike near the horizon and spacelike near infinity for any $\Omega \ne 0$. So there is always a circular geodesic which is invariant under the Killing field, and putting a black hole on this orbit could generate radiation which is invariant under $\xi$.  But the corresponding solution is not asymptotically flat as the radiation both to and from infinity is periodic in time.  Instead of generating this solution, starting the binary on a similar orbit with asymptotically flat boundary conditions must lead to coalescence.  Clearly, the AdS boundary conditions play a crucial role in the above construction.

%\section{Implications for gauge/gravity duality}

\section{Discussion}

We have argued that many solutions asymptotic to global AdS are nonlinearly stable, despite the fact that global AdS itself is unstable. The key difference is that normal modes in global AdS have frequencies which are all integer multiples of the AdS frequency, so there are infinite towers of resonances in the perturbation expansion. In contrast, other solutions with less symmetry do not have exact resonances. While approximate resonances (associated with high angular momentum modes) still exist, we have seen that they are not strong enough to cause an instability. Our arguments were based on formal perturbation theory, though the qualitative similarity between our results and established theorems for the nonlinear Schr\"odinger equation encourages the belief that they can be placed on a more rigorous footing. There is also numerical support for the nonlinear stability of boson stars \cite{Liebling}. The suggestion \cite{Dias:2011ss} that there may be a  theorem showing generic solutions asymptotic to global AdS are singular is clearly incorrect\footnote{See \cite{Ishibashi} for some of the difficulties in trying to prove such a theorem.}.  While our methods allow us to study only the stability of stationary backgrounds, we expect perturbations of oscillon solutions to behave similarly.

Particular solutions may of course have excitations with additional exact or approximate resonances at finite frequency.  But as discussed in section \ref{NLSE},
nonlinear instabilities typically require an infinite tower of (perhaps approximate) resonances.  Since excitations of backgrounds asymptotic to global AdS have a discrete spectrum, any such tower involves modes of arbitrarily high frequency so that our analysis should suffice\footnote{Solutions asymptotic to Poincar\'e AdS can also feature what is effectively an infinite tower of resonances at low frequency.  In particular, at leading order in long wavelength along the AdS boundary, low frequency perturbations of planar AdS black holes are described \cite{Bhattacharyya:2008jc} by the Euler equations for a conformal fluid on the boundary spacetime.  These equations have exact zero modes associated with divergence-free fluid flows.  It is natural to expect such zero modes to lead to nonlinear instabilities of planar AdS black holes.  While black holes asymptotic to global AdS should be nonlinearly stable for small enough perturbations, fixing the size of the perturbation while taking the limit of a large such black hole should result in behavior much like that of the nonlinearly unstable planar case.  We believe this explains, from the gravitational side,  the results of \cite{Lehner}.}.

This argues that various ``excited states" such as geons and boson stars can be stable. However, one expects to also construct stable ground states by modifying the boundary conditions at infinity so the lowest energy solution differs from AdS. This includes putting a metric on the sphere at infinity which is not conformally flat. In theories with scalar fields whose mass is close to the Breitenl\"ohner-Freedman bound, one has a large class of allowed boundary conditions for the scalar field. In many cases, the ground state includes a nonzero scalar field \cite{Amsel:2007im}. Finally, the AdS soliton \cite{Witten:1998zw} is believed to be the lowest energy solution for its boundary conditions.  All of these solutions are likely to be nonlinearly stable by the arguments given earlier. In fact, they fall into the category of solutions mentioned in footnote \ref{sources} which should be even more stable than the geons, boson stars, and black holes addressed in the main text.

The picture that emerges is that AdS is very special. Its high degree of symmetry leads to a large number of resonances and causes its nonlinear instability. Most other solutions do not have this feature.

We now consider the implications of these results for gauge/gravity duality. It was argued in \cite{Dias:2011ss} that  the instability of AdS has a simple interpretation in the dual field theory: Since black holes have the maximum entropy in the classical limit, the instability toward forming small black holes can be viewed as thermalization (in a microcanonical ensemble since the energy is fixed). The existence of geons was puzzling from the dual field theory viewpoint, since they represent finite energy excitations of the strongly coupled large $N$ field theory which never thermalize. We now see that AdS is very special. Not only are geons stable, but so are most nontrivial ground states. This would seem to contradict the idea that energy added  to a system generically thermalizes. Note that since we have finite backreaction in the bulk, the energy we add is always $O(N^2)$ and is not really small. So this lack of thermalization is not just a property of a few low lying energy eigenstates. However, one must remember that these results only hold at large $N$. In many ways,  large $N$ field theories act classically, and classical field theories almost never thermalize. If there is a potential landscape, a small amount of energy added near one local minimum will stay near that minimum and never explore other configurations. From this viewpoint, pure AdS is like having no potential.

At finite $N$, one expects quantum tunneling to cause the states of the boundary theory (on $S^n\times {\mathbb R} $) to explore all accessible regions, so many operators will behave thermally. Similarly, in the bulk we expect added energy to generically thermalize, (although  the maximum entropy configuration can now be a gas of radiation and not a black hole).

\vskip 1cm
\centerline{\bf Acknowledgements}
\vskip .5cm
It is a pleasure to thank P. Bizon, M. Dafermos,  L. Lehner, and I. Rodnianski  for discussions. This work was supported in part by NSF grants PHY-0855415 and PHY-1205500. We thank the organizers and participants of the ``Exploring AdS/CFT Dualities in Dynamical Settings", Perimeter Institute Workshop (2012) for stimulating discussions. OD thanks the Yukawa Institute for Theoretical Physics (YITP) at Kyoto University, where part of this work was completed during the YITP-T-11-08 programme  ``Recent advances in numerical and analytical methods for black hole dynamics".

\end{document}